\documentclass{aa}  

\usepackage{graphicx}
\usepackage{tikz}
\usetikzlibrary{arrows.meta}
\usetikzlibrary{ decorations.pathmorphing, decorations.pathreplacing, decorations.shapes}
\usepackage{tcolorbox}
\newtcolorbox{mybox}[1]{colback=blue!5!white,colframe=blue!75!black,fonttitle=\bfseries,title=#1}
\usepgflibrary{shadings}
\usepackage{xcolor}
\usepackage[colorlinks=true,allcolors=blue]{hyperref}

\usepackage{txfonts}
\usepackage{bigstrut}
\begin{document} 
\definecolor{sunyellow}{RGB}{255,223,0}
\definecolor{lightblue}{RGB}{173,216,230}

\title{Modeling of the Yarkovsky effect}
\subtitle{I. Thermal dependence}
\titlerunning{Modeling of the Yarkovsky effect. I. Thermal dependence}

\author{O. Golubov\inst{1}, O. Mikhalchenko\inst{1}, V. Lipatova\inst{2}, D. J. Scheeres\inst{3}}

\institute{Institute of Astronomy, V. N. Karazin Kharkiv National University, 4 Svobody sq., Kharkiv 61022, Ukraine
\and
Zentrum f\"{u}r Astronomie, Institut f\"{u}r Theoretische Astrophysik, Universit\"{a}t Heidelberg, Albert-Ueberle-Str. 2, D-69120 Heidelberg, Germany
\and
Smead Department of Aerospace Engineering Sciences, University of Colorado Boulder, Boulder, CO, USA \\
\email{oleksiy.golubov@karazin.ua}
}

\date{To be submitted to Astronomy and Astrophysics}

% \abstract{}{}{}{}{} 
% 5 {} token are mandatory
 
\abstract
{The Yarkovsky effect is an important non-gravitational torque experienced by asteroids. It results from the thermal inertia of the asteroid soil, and thus its computation requires thermal modeling of the asteroid.}
{Here, we investigate how the Yarkovsky effect created by one flat surface element on the asteroid depends on the thermal parameter. Then we integrate the Yarkovsky force over the surface of a spherical asteroid. In the next articles of the series, we will investigate the Yarkovsky force for bodies of more complicated shape and compute the change of the asteroid's orbit.}
{Using perturbation theory, we analytically solve the one-dimensional heat conduction equation under the surface element and use the solution to analytically compute the Yarkovsky effect in the limiting cases of big and small thermal parameters. Then we analytically sew the two solutions to obtain an approximate expression for the Yarkovsky effect valid for both large and small thermal parameters. We verify it by comparing it to the results of the numerical solution of the heat conduction equation. Finally, we fit the parameters in an approximate analytical expression to improve the agreement with the numerical results.}
{We obtain a fully analytical expression for the Yarkovsky effect of one surface element, which is composed of 4 terms and is precise for any value of the thermal parameter up to the accuracy of 15\%. We also propose a fitted expression of the same form, which is precise up to 0.15\%. They both much supersede the traditionally used 3-parameter fit, which produces an error up to 50\% and, in particular, is about a factor of 2 wrong for small values of thermal parameters. Finally, we apply the new formalism to evaluate an analytical expression for the Yarkovsky force acting on a spherical asteroid, ending up in a 0.1{\%}-accurate fitted expression for an asteroid with 0 obliquity on a circular orbit and a less accurate expression for the general case.}
{}


\maketitle
%
%-------------------------------------------------------------------

\section{Introduction}
When an asteroid rotates around its axis, the maximal illumination at each point is attained at midday. However as the surface material has a certain thermal inertia, it takes some time for the thermal energy to accumulate in it, so the maximum temperature is reached later in the afternoon. As a result, the evening hemisphere of the asteroid is on average warmer than its morning hemisphere, and thus emits more thermal infrared radiation. The resulting difference between the light pressure recoil forces on the two hemispheres pushes the asteroid along the orbital velocity for a prograde rotator and against the orbital velocity for a retrograde rotator, secularly changing its orbit \citep{opik1951, bottke2002astIII, vokrouhlicky2015a}.

This so-called Yarkovsky effect is crucially important for the orbital evolution of the Solar System small bodies by spreading asteroid families into prominent Yarkovsky \textit{V-shapes} and driving the main-belt asteroids to the Kirkwood gaps, and thus re-supplying the population of near-Earth objects \citep{vokrouhlicky2015a}. Computation of the Yarkovsky effect is also of a great practical importance for the risk assessment of potentially hazardous objects, as this non-gravitational force often constitutes the major source of uncertainty for the asteroid orbit prediction \citep{farnocchia_etal2015}. Another practical application of the Yarkovsky effect, and a more optimistic one, is using it to evaluate asteroid densities from known thermal inertias or evaluate thermal inertias from assumed densities \citep{dziadura_etal2023, novakovic_etal2024}. 

The theory of the Yarkovsky effect was gradually established throughout the second half of the 20th century.
In our opinion, there are four articles that can be considered ``pioneering'' in the study of this effect, namely: \cite{opik1951}, \cite{radzievskii1952}, \cite{peterson1976a} and \cite{rubincam1995}. 
Ernst J. {\"O}pik \citep{opik1951} was the first to name, describe and briefly investigate the drag that occurs when there is an asymmetry in the re-emission of absorbed solar radiation by the surface of a very small spinning body orbiting the Sun. {\"O}pic himself pointed to Ivan O. Yarkovsky as the discoverer of the effect, referring only from memory to 
a certain pamphlet written by Yarkovsky and published in 1901 that was found in the archives only more than 50 years after {\"O}pic’s reference to it (see \cite{beekman2006}; the original \cite{yarkovsky1901} pamphlet is available as an appendix to \cite{broz2006}). It is a pity that in {\"O}pic’s paper, the study of the Yarkovsky effect is really very brief, without a detailed description of the used assumptions and intermediate calculations. Around the same time, independently, Vladimir V. Radzievskii \citep{radzievskii1952} studied a similar effect, but unlike {\"O}pic, in his study, Radzievskii described in detail a simple analytical solution of the heat equation and explored the range of its applicability in estimating the magnitude of the emerging effect.

Charles A. Peterson \citep{peterson1976a}, having analyzed the cosmic ray exposure ages of meteorites, apparently was the first to propose a two-stage mechanism for delivering small asteroid fragments from the main belt into an orbit crossing that of Earth: the orbit of the fragment changes slowly under the influence of the Yarkovsky effect until the body is caught by one of the gravitational resonances with planets, which accelerates its further dynamical evolution. The author emphasized the importance of the selectivity of the Yarkovsky effect, whose influence on asteroids rapidly decreases with increasing asteroid size, while gravitational perturbations accelerate all asteroids equally. Having performed calculations very similar to those given in Radzievskii's paper, Peterson was nevertheless the first to discover that the Yarkovsky force reaches a maximum value at a certain asteroid spin rate, but dramatically decreases for very fast or very slow rotators.

After a long pause in studies of the dynamical evolution of small bodies under the influence of the Yarkovsky force, an article by David P. Rubincam appeared \citep{rubincam1995}, in which he described and modeled in detail a component of the Yarkovsky effect overlooked by the previous authors: the ``seasonal'' effect due to temperature variation during asteroid's revolution around the Sun. It was his idea to separately consider the ``diurnal'' and ``seasonal'' Yarkovsky effects, which depend, respectively, on the spin rate of the asteroid and on its orbital mean motion.

By the late 1990-s, the importance of the Yarkovsky effect in explaining many aspects of the asteroid population had been realized \citep{farinella1998}. 
%This urged an excessive study of the Yarkovsky effect as a function of thermal properties, asteroid shape, orbit, and spin state. 
This urged an excessive study of the Yarkovsky effect as a function of thermal properties, asteroid albedo, shape, orbit, and spin state, and also as a superposition of diurnal and seasonal components.
Such an investigation was conducted by \citeauthor{vokrouhlicky1998a} (\citeyear{vokrouhlicky1998a,vokrouhlicky1998b,vokrouhlicky1999}), \citeauthor{vokrouhlicky_farinella1998} (\citeyear{vokrouhlicky_farinella1998, vokrouhlicky_farinella1999}), \cite{vokrouhlicky_broz1999}, \cite{vokrouhlicky_bottke2001}, and \cite{vokrouhlicky2006}. As a result, the authors reached a conclusion that many of the listed complications changed the value of the Yarkovsky effect by a margin, small by the standards of 1990-s, and ``put on hold'' further analysis of some of these additional factors (e.g., asteroid shape, surface roughness, or the complex interplay between diurnal and seasonal effects).

%As a result, the authors reached a conclusion that many of the listed complications changed the value of the Yarkovsky effect by a margin, small by the standards of 1990-s, and excluded some of the additional factors from further analysis (such as the asteroid shape and the orbit eccentricity).

Thus, by ignoring the parameters thought to be of lesser importance, a model of the Yarkovsky effect was formulated, which can be called the ``standard'' one \citep{vokrouhlicky2001,bottke2002astIII,vokrouhlicky2015a}.
The main feature of this model is the analytical representation of the solution, achieved by using a Taylor series expansion to linearize the Stefan--Boltzmann law, thereby avoiding the much more complex heat conduction problem with a non-linear boundary condition.
%The analysis of the Yarkovsky effect has generally relying on the classical computation of the Yarkovsky effect when simple analytical results are needed, while taken a numerical solution approach for precise computations of the thermal response. 
The resulting linearized analytical theory is easily incorporated into different N-body integrators used for simulations of the long-term orbit evolution of asteroids \citep{bottke2000, grimm_etal_2022, fenucci_novakovic2022, ferich_etal_2022}.

In some cases, numerical modeling can be used instead of analytical theory to accurately calculate the thermal response when computing the Yarkovsky effect \citep{spitale_greenberg2001,rozitis_green2012}. When in 2003 the Yarkovsky effect was first unambiguously confirmed by observations of asteroid (6489) Golevka, a numerical computation of the Yarkovsky force was proved to be in very good agreement with the observations \citep{chesley2003}. For that computation, the detailed radar shape model of Golevka was used in conjunction with a numerical heat transfer solution to determine the changes in the surface temperature of the asteroid during its complete orbital revolution. 
A similar numerically intensive approach was later used to successfully predict the density of the asteroid (101955) Bennu, based on measuring the orbital drift of that body and fitting it to precise Yarkovsky computations \citep{chesley2014,scheeres2019}. 

%It should be noted that although such a numerical solution of the heat conduction equation is highly accurate
%It should be noted that although with good numerical solution of the heat conduction equation one can go without theoretical modeling at all,
%Certainly, with good numerical simulations one can go without theoretical modeling at all,
%It should be noted that although numerical solution of the heat conduction equation is highly accurate, it is nevertheless still be very expensive in terms of time and resources spent. 

Nevertheless, the numerical solution of the heat equation in the context of the Yarkovsky effect theory remains quite expensive in terms of time and resources required. 
Furthermore, conducting such precise simulations for typically very uncertain data is not only unnecessary but even harmful, as such a ``black-box'' numerical approach conceals the physics of the effect. An analytical model of the Yarkovsky effect can be preferred in many cases, as it is clearer, better controlled, and just as accurate as a fully numerical model except for the few most studied asteroids.

The main problem is that the existing analytical model of the Yarkovsky effect is subject to severe limitations inherited from the underlying linearized thermal model. The linearization, based on the Taylor approximation of the term $T^4$, remains valid as long as the surface temperature deviates only slightly from its mean value. This assumption held when the Yarkovsky effect theory was first formulated for dust particles, meteoroids, and monolithic asteroidal fragments up to several meters in size \citep{opik1951, radzievskii1952, peterson1976a, burns_etal1979, vokrouhlicky1999}, whose rotation is so fast that the temperature indeed cannot change much during their rotation period. However, when the same theory was later applied to asteroids with much slower axial rotation \citep{farinella_vokrouhlicky1999,bottke_etal2001}, the applicability conditions of the linear model got stretched. At first, this deficiency of the analytical model was explicitly mentioned by the researchers \citep{radzievskii1952,peterson1976a,spitale2001}, but in later works such clarifications were omitted.

As a result, presently the ``standard'' linear theory is routinely applied to asteroids whose diurnal and annual temperature variations are of the same order as the temperature itself, putting it on the edge or outside the area of applicability of the linear model.
Furthermore, in the light of surging interest in ultra-rapid rotators, \cite{marceta_etal2025} misinterpret the limitations of this model, devoting significant effort to evaluating its validity for super-fast rotating asteroids, to which it is applicable by design.

In fairness, efforts are consistently made to balance the efficiency of low-accuracy analytical calculations with the precision of computationally expensive numerical simulations through the use of hybrid iterative techniques; yet such approaches address primarily the seasonal Yarkovsky force, sacrificing the diurnal component (e.g., see \cite{vokrouhlicky_farinella1999}, \cite{paoli_etal2026}).

%The recent attempts to construct a non-linear model for the seasonal Yarkovsky effect \citep{paoli_etal2026} still reside on uncertain models for diurnal temperature averaging. An exception arises in the case of annual surface-temperature variations on asteroids in highly eccentric orbits, where efforts to develop analytical models that avoid the limitations of linearization remain an active area of research. While neglecting diurnal variations and preserving the linearization scheme, \cite{paoli_etal2026} abandoned the assumption of a constant average temperature. Instead, its value was allowed to vary with time and radial distance from the asteroids`s center. Authors noted that their new approach appears to be adequate for orbits up to eccentricity $e \simeq 0.4$. However, it remains unclear whether such refinements outweigh the significant errors inherent in the ``standard'' linear theory.

Therefore, the development of a more complete and accurate analytical model presents a highly relevant area for investigation. A new comprehensive theory is needed to supersede the insufficient classical approach to the modeling of the Yarkovsky effect and to match the abundance of new data on asteroid shapes and thermal properties, as well as the ever-growing accuracy of the Yarkovsky drift rate detection.
With this article, we start a series of papers devoted to the elaboration of the theory of the Yarkovsky effect. This particular article will set the framework of our analysis and then focus on the dependence of the Yarkovsky pressure on the thermal model of the asteroid, whereas the following articles will consider in more detail the impact of the asteroid's shape, orbit, and other parameters. 

In the current work, we mostly focus on an asteroid of a spherical shape with zero obliquity on a circular orbit around the Sun and come up with an updated analytical expression for its Yarkovsky force. Before starting, we discuss in Section \ref{sec:scope} the general scope of the problem of the Yarkovsky drift and introduce the formalism that will be used in the following articles of the series. In Section \ref{sec:one_el_equator}, we analytically consider the thermal model of an element on the asteroid's equator and verify our analytical results by numerical simulations. In Section \ref{sec:latitude_dependence}, we generalize these results for arbitrary latitude on the asteroid, following closely in parallel to the approach from Section \ref{sec:one_el_equator}. Then, in Section \ref{sec:spherical_ast_integr}, we integrate the Yarkovsky pressure over the surface of the spherical asteroid. In Section \ref{sec:results}, we summarize obtained results. In Section \ref{sec:discussion}, we discuss the implications of our results, which, despite the simplicity of the assumed model, substantially supersede the accuracy of the classical approach.

\section{Scope of the problem}
\label{sec:scope}
Let us express the Yarkovsky drift rate in terms of several functions that will be determined later in this article and in the subsequent articles of the series.

\begin{figure*}
\begin{center}
\begin{tikzpicture}[scale=0.6] 
    
    \begin{scope}[shift={(-8,0)}]
        \draw[thick] (0,0) circle (5cm);
        
        \draw[thick, gray] (0:3.5) arc[start angle=0, end angle=-180, radius=3.5cm];
        \draw[thick,-latex,gray] (-1:3.5cm) -- (1:3.5cm);
        
        \fill[sunyellow] (0,0) circle (0.8cm);
        \foreach \angle in {0,15,...,345} {
            \draw[sunyellow, fill=sunyellow] (\angle:0.8cm) -- ++(\angle+7.5:0.5cm) -- ++(\angle+172.5:0.5cm) -- cycle;
        }
        
        \foreach \angle in {0, 45, 90, 135, 180, 225, 270, 315} {
            \draw[very thick, -latex , teal] (\angle:5cm) -- (\angle+15:6.5cm);
            \shade[left color=lightblue!100, right color=orange!50,shading angle=\angle-45] (\angle:5cm)  circle (0.9cm);
            \fill[black] (\angle:5cm) circle (0.1cm);
            
            \draw[gray] (90:5cm) +(-90:0.5cm) arc[start angle=-90, end angle=90, radius=0.5cm];
            \draw[thick,-latex,gray] (90:5.5cm) -- (93:5.6cm);
        }
    \end{scope}
    
    \begin{scope}[shift={(6,0)}] 
        \draw[thick] (0,0) circle (5cm);
        
        \draw[thick, gray] (0:3.5) arc[start angle=0, end angle=-180, radius=3.5cm];
        \draw[thick,-latex,gray] (-1:3.5cm) -- (1:3.5cm);
        
        \fill[sunyellow] (0,0) circle (0.8cm);
        \foreach \angle in {0,15,...,345} {
            \draw[sunyellow, fill=sunyellow] (\angle:0.8cm) -- ++(\angle+7.5:0.5cm) -- ++(\angle+172.5:0.5cm) -- cycle;
        }
        \draw[very thick, -latex , teal] (0:5.1cm) -- (-20:5.355cm);
        \shade[right color=lightblue!100, left color=orange!50,shading angle=0] (0:5cm)  circle (0.9cm); 
        \draw[black] (0:4.1cm) -- (0:5.9cm);
        \node at (5,0.3) {$N$};
        \node at (5,-0.3) {$S$};
        
        \shade[left color=lightblue!65!orange, right color=lightblue!65!orange,shading angle=0] (45:5cm)  circle (0.9cm); 
        \draw[black] (39:5.7cm) -- (53.5:4.45cm);
        \node at (3.54,3.9) {$N$};
        \node at (3.54,3.3) {$S$};

        \draw[very thick, -latex , teal] (90:5cm) -- (90:6.7cm);
        \draw[thick] (0,3.5) ellipse (0.6cm and 0.2cm);
        \fill[white] (0,3.7) circle (0.4cm);
        \draw[-latex] (0.4,3.65) -- (0.2,3.75);
        \draw[thick, -latex , gray] (90:3cm) -- (90:5cm);
        \shade[left color=lightblue!100, right color=orange!50,shading angle=0] (90:5cm)  circle (0.9cm); 
        \draw[black] (80:5.1cm) -- (100:5.1cm);
        \node at (0,5.3) {$N$};
        \node at (0,4.7) {$S$};

        \draw[very thick, -latex , teal] (135:5cm) -- (135-14:7cm);
        \shade[left color=lightblue!120!orange, right color=orange!100!lightblue,shading angle=0] (135:5cm)  circle (0.9cm); 
        \draw[black] (141:5.7cm) -- (126.5:4.45cm);
        \node at (-3.54,3.9) {$N$};
        \node at (-3.54,3.3) {$S$};

        \draw[very thick, -latex , teal] (180:5.1cm) -- (180-20:5.355cm);
        \shade[left color=lightblue!100, right color=orange!50,shading angle=0] (180:5cm)  circle (0.9cm); 
        \draw[black] (180:4.1cm) -- (180:5.9cm);
        \node at (-5,0.3) {$N$};
        \node at (-5,-0.3) {$S$};
        
        \shade[left color=lightblue!65!orange, right color=lightblue!65!orange,shading angle=0] (225:5cm)  circle (0.9cm);
        \draw[black] (219:5.7cm) -- (233.5:4.45cm);
        \node at (-3.54,-3.9) {$S$};
        \node at (-3.54,-3.3) {$N$};

        \draw[very thick, -latex , teal] (270:5cm) -- (270:6.7cm);
        \shade[right color=lightblue!100, left color=orange!50,shading angle=0] (270:5cm)  circle (0.9cm); 
        \draw[black] (280:5.1cm) -- (260:5.1cm);
        \node at (0,-5.35) {$S$};
        \node at (0,-4.7) {$N$};

        \draw[very thick, -latex , teal] (315:5cm) -- (315-14:7cm);
        \shade[right color=lightblue!120!orange, left color=orange!100!lightblue,shading angle=0] (315:5cm)  circle (0.9cm); 
        \draw[black] (321:5.7cm) -- (306.5:4.45cm);
        \node at (3.54,-3.9) {$S$};
        \node at (3.54,-3.3) {$N$};
        
    \end{scope}
\end{tikzpicture}
\caption{An illustration of the diurnal (left panel) and seasonal (right panel) Yarkovsky effect}
\label{fig:Yarkovsky-illustration}   
\end{center}
\end{figure*}
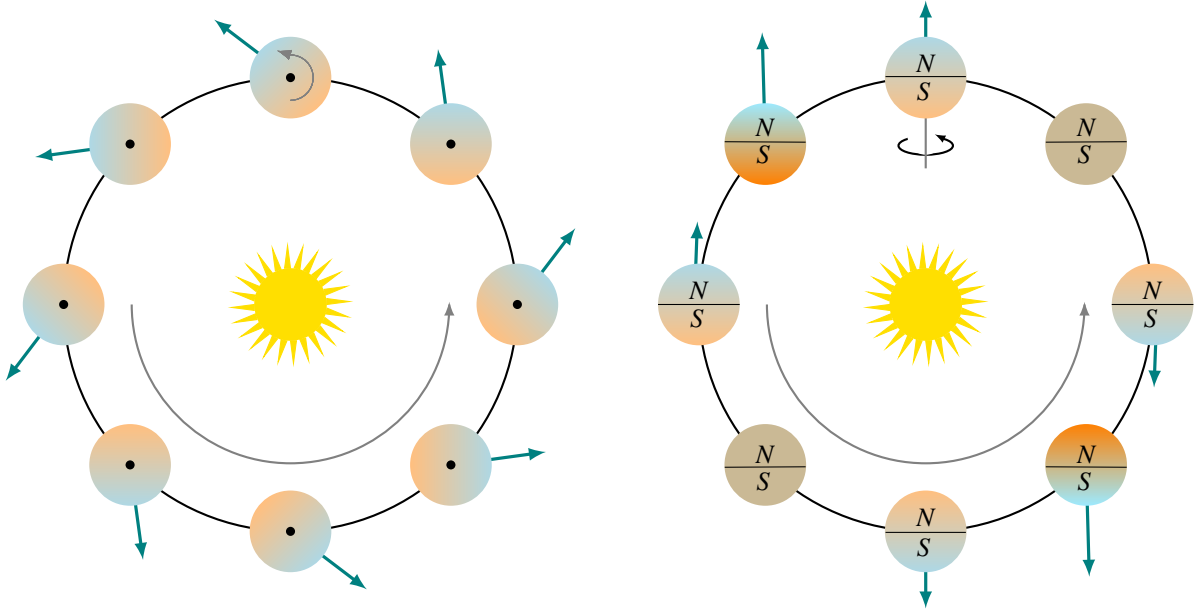

Following tradition, we subdivide the Yarkovsky effect into the diurnal and seasonal components, which are explained in Figure \ref{fig:Yarkovsky-illustration}.

\subsection{Toy model for the diurnal Yarkovsky effect on a spherical asteroid with zero obliquity on a circular orbit}
We start with the simplest toy model, namely a spherical asteroid with zero obliquity on a circular orbit. From the solar power flux $\Phi$, the asteroid absorbs the fraction $1-A$, with $A$ being its bolometric Bond albedo. Then it re-emits this energy and experiences the recoil pressure of the order of $(1-A)\Phi/c$, with $c$ being the speed of light. As the asteroid rotates, the pressure force continuously changes its direction, thus its time-averaged component in the direction of the asteroid's orbit is proportional to but smaller than $(1-A)\Phi/c$, with the proportionality coefficient depending on the latitude and the thermal properties of the surface. Simple thermal models are fully defined by one single parameter $\theta$, called thermal parameter, which will be discussed in more detail in Section \ref{sec:one_el_equator}. Therefore, we write that the mean Yarkovsky force acting upon a unit surface element on the asteroid's equator is equal to
\begin{equation}
\frac{dF}{dS}=\frac{(1-A)\Phi p_\theta(\theta)}{c},
\label{p_theta_def1}
\end{equation}
where $p_\theta$ is some function determined by the thermal model. Then we expect that for a surface element on a latitude $\psi$ the force can be approximately obtained via multiplication by $\cos^2\psi$, with one $\cos\psi$ describing the projection of the solar rays onto the surface element, and the other $\cos\psi$ describing the projection of the normal light-pressure force onto the asteroid's orbit. By integrating these forces over the surface of the spherical asteroid with the radius $R$, we get the Yarkovsky force
\begin{align}
F&=\frac{(1-A)\Phi}{c}p_\theta(\theta)\int\limits_{-\pi/2}^{\pi/2}2\pi R^2\cos(\psi)d\psi\cos^2(\psi)=\nonumber\\
&=\frac{(1-A)\Phi R^2}{c}p_\theta(\theta)\frac{8\pi}{3}.
\label{FYark-initial}
\end{align}
Here, $2\pi R^2\cos(\psi)d\psi$ is the surface area encompassed between the latitudes $\psi$ and $\psi+d\psi$.

An asteroid on a circular orbit has the angular momentum $L=m\sqrt{M_\odot Gr}$, where $r$ is the heliocentric distance, $M_\odot$ is the mass of the Sun and $m=\frac{4}{3}\pi R^3\rho$ is the mass of the asteroid, with $\rho$ being its density. By differentiating $L$ and equating its derivative to the torque created by the Yarkovsky force $F$ with the lever arm $r$, we get
\begin{equation}
\frac{4}{3}\pi R^3\rho\cdot\frac{1}{2}\sqrt{\frac{M_\odot G}{r}}\Dot{r}=\frac{(1-A)\Phi}{c}p_\theta(\theta)\frac{8\pi}{3}r.
\end{equation}

From this equation, we derive $\Dot{r}$. By the way, we get rid of $\Phi$ and express it in terms of the solar luminosity $L_\odot$ as $\Phi={L_\odot}/(4\pi r^2)$.

\begin{equation}
\Dot{r}=\frac{(1-A)L_{\mathrm{\odot}}}{\pi Rc\rho\sqrt{M_{\mathrm{\odot}}Gr}}p_\theta(\theta).
\end{equation}

\subsection{Toy model for the seasonal Yarkovsky effect on a spherical asteroid with the obliquity 90$^\circ$ on a circular orbit}

Consider the next toy model, now for the seasonal Yarkovsky effect for a spherical asteroid with the obliquity 90$^\circ$ on a circular orbit, as shown in the right panel of Figure \ref{fig:Yarkovsky-illustration}. As seen from the Sun, the asteroid's polar region uniformly rotates in the orbital plane with the period of one asteroid year, just as an equatorial surface element in the left panel of Figure \ref{fig:Yarkovsky-illustration}, but in the retrograde direction. Therefore, similarly to the previous subsection, it creates the mean Yarkovsky force along the orbit equal to $-(1-A)\Phi p_\theta(\theta_s)/c$, with the minus sign standing for the retrograde rotation, $\theta_s$ being the seasonal thermal parameter, and $p_\theta$ being the same function as before.

The surface elements on other latitudes experience different Yarkovsky pressures. As the pressure must be maximal at the poles ($\psi=\pm\pi/2$), zero at the equator ($\psi=0$), and smooth around the sphere, we can expect it to be approximately proportional to $\sin^2\psi$. Now, integrating the seasonal Yarkovsky force over the spherical asteroid, we obtain
\begin{align}
F&=-\frac{(1-A)\Phi}{c}p_\theta(\theta_s)\int\limits_{-\pi/2}^{\pi/2}2\pi R^2\cos(\psi)d\psi\sin^2(\psi)=\nonumber\\
&=-\frac{(1-A)\Phi R^2}{c}p_\theta(\theta_s)\frac{4\pi}{3}.
\label{FYark-initial-seasonal}
\end{align}
This result differs from Eq. (\ref{FYark-initial}) by the different sign, the factor 2 in the denominator, and $\theta_s$ standing in the place of $\theta$. These differences make it all the way to the final expression for the seasonal Yarkovsky orbital drift:
\begin{equation}
\Dot{r}=-\frac{(1-A)L_{\mathrm{\odot}}}{2\pi Rc\rho\sqrt{M_{\mathrm{\odot}}Gr}}p_\theta(\theta_s).
\end{equation}

\subsection{Simplified consideration of eccentricity}
Now consider a more complicated case of the diurnal Yarkovsky effect for an asteroid on an elliptical orbit with the eccentricity $e$ and the semimajor axis $a$. We apply the same equation (\ref{FYark-initial}) for the Yarkovsky force, taking into account that the solar constant at distance $r$ scales as $\Phi={L_\odot}/(4\pi r^2)$, but neglecting for the time being that $p(\theta)$ can also change with $r$. The time-averaged power of the Yarkovsky force equals to
\begin{equation}
\langle P\rangle=\langle \Vec{F}\cdot\Vec{v}\rangle=\langle F v_\perp\rangle=\frac{2(1-A) R^2 L_\odot}{3c}p_\theta(\theta)\left\langle \frac{v_\perp}{r^2}\right\rangle,
\label{F-dot-v}
\end{equation}
where $v_\perp$ is the velocity component of the asteroid in the direction perpendicular to the Sun. We express $v_\perp=r\frac{d\upsilon}{dt}$ in terms of the asteroid's true anomaly $\upsilon$ and re-write the time-averaging as the time integral over the orbital period $P_\mathrm{orb}=2\pi\sqrt{a^3/(GM_\odot)}$ divided by $P_\mathrm{orb}$,
\begin{align}
\langle P\rangle&=\frac{2(1-A) R^2 L_\odot}{3cP_\mathrm{orb}}p_\theta(\theta)\int\limits_0^{P_\mathrm{orb}}\frac{1}{r^2} r\frac{d\upsilon}{dt}dt=\nonumber\\
&=\frac{2(1-A) R^2 L_\odot}{3c\cdot P_\mathrm{orb}}p_\theta(\theta)\int\limits_0^{2\pi}\frac{1}{r} d\upsilon.
\label{P_via_upsilon_integr}
\end{align}
The radius $r$ is connected with the true anomaly $\upsilon$ by the equation of ellipse in polar coordinates:
\begin{equation}
r=\frac{(1-e^2)a}{1+e\cos\upsilon}.
\label{ellipse_polar}
\end{equation}
Substituting Eq. (\ref{ellipse_polar}) into Eq. (\ref{P_via_upsilon_integr}), we obtain
\begin{align}
\langle P\rangle&=\frac{2(1-A) R^2 L_\odot}{3cP_\mathrm{orb}}p_\theta(\theta)\int\limits_0^{2\pi}\frac{1+e\cos\upsilon}{(1-e^2)a} d\upsilon=\nonumber\\
&=\frac{2(1-A) R^2 L_\odot}{3c\cdot 2\pi\sqrt{a^3/(GM_\odot)}}p_\theta(\theta)\frac{2\pi}{(1-e^2)a}
\label{P_via_upsilon}
\end{align}
For a Keplerian orbit the total energy $E_\mathrm{orb}$ is connected with the semimajor axis by the equation
\begin{equation}
E_\mathrm{orb}=-\frac{GM_\odot m}{2a} \Rightarrow \Dot{E}_\mathrm{orb}=\frac{GM_\odot m}{2a^2}\Dot{a}.
\end{equation}
By equating the energy derivative to the mean power of the Yarkovsky force $\langle P\rangle$, we derive the following expression for $\Dot{a}$:
\begin{equation}
\Dot{a}=\frac{(1-A)L_{\mathrm{\odot}}}{\pi Rc\rho\sqrt{M_{\mathrm{\odot}}Ga}}p_\theta(\theta)\frac{1}{1-e^2}.
\label{e-diurnal}
\end{equation}

Concerning the seasonal component, we can in the first approximation, expect that Eq. (\ref{FYark-initial-seasonal}) can be substituted into Eq. (\ref{F-dot-v}) instead of Eq. (\ref{FYark-initial}) and followed all the way through, thus resulting into
\begin{equation}
\Dot{a}=-\frac{(1-A)L_{\mathrm{\odot}}}{2\pi Rc\rho\sqrt{M_{\mathrm{\odot}}Ga}}p_\theta(\theta_s)\frac{1}{1-e^2}.
\label{e-seasonal}
\end{equation}

\subsection{Simplified consideration of obliquity and the seasonal Yarkovsky effect}
\begin{figure}
\centering
\begin{tikzpicture}
\draw[-{Latex[length=3mm, width=2mm]},very thick,red] (0,0) -- (0,2.667);
\draw[-{Latex[length=3mm, width=2mm]},very thick,teal!70!lime] (0,0) -- (1.887,1.887);

\draw (0,2) arc (90:45:2);
\draw[-{Latex[length=3mm, width=2mm]}] (1.21,1.59) -- (1.41,1.41);

\draw (0,1.47) arc (90:405:1.47);
\draw[-{Latex[length=3mm, width=2mm]}] (1.24,0.804) -- (1.04,1.04);

\draw (0,1.) arc (90:-90:0.9);
\draw (0,-0.8) arc (270:90:0.7);
\draw (0,0.6) arc (90:45:0.6);
\draw[-{Latex[length=3mm, width=2mm]}] (0.22,0.55) -- (0.4242,0.4242);

\node[red] at (-0.45,2.4) {$\mathbf{L}_\mathrm{orb}$};
\node[teal!70!lime] at (2.2,1.55) {$\mathbf{L}_\mathrm{spin}$};
\node[] at (0.8,2.05) {$\varepsilon$};
\node[] at (0,-0.5) {$\pi+\varepsilon$};
\node[] at (0,-1.22) {$\pi-\varepsilon$};

\end{tikzpicture}
\caption{Illustration of the symmetry properties of the obliquity $\varepsilon$.}
\label{fig:varepsilon-symmetry}
\end{figure}
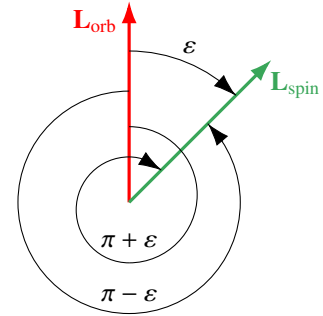

In the previous three subsections, we considered asteroids with obliquity $\varepsilon=0$ (diurnal effect) or $\varepsilon=\pi/2$ (seasonal effect), whereas in the general case the obliquity belongs to the range $0<\varepsilon<\pi$. The obliquity is defined as the angle between the angular momenta of the asteroid's spin and orbital motion. The symmetry properties of the Yarkovsky effect as a function of $\varepsilon$ is easier to see if we extend the definition of $\varepsilon$ to the entire number line $-\infty<\varepsilon<\infty$, but keep in mind the equivalence between $\varepsilon$, $2\pi-\varepsilon$, and $2\pi+\varepsilon$ (see Figure \ref{fig:varepsilon-symmetry}).

From the periodicity of the Yarkovsky force $F$ as a function of $\varepsilon$ with the period $2\pi$, it follows that it can be decomposed into the Fourier series,
\begin{equation}
F(\varepsilon)=\sum\limits_{n=0}^\infty (a^\varepsilon_n\cos{n\varepsilon}+b^\varepsilon_n\sin{n\varepsilon}),
\label{eps-series}
\end{equation}
where $a^\varepsilon_n$ and $b^\varepsilon_n$ are the Fourier coefficients.

This series must satisfy the condition $F(2\pi-\varepsilon)=F(\varepsilon)$. The substitution of $2\pi-\varepsilon$ instead of $\varepsilon$ in Eq. (\ref{eps-series}) leaves all the cosines intact, but changes the signs of all the sines. The series can remain unaffected by such change, only if all the coefficients in front of sine terms are zeros, $b^\varepsilon_n=0$, thus
\begin{equation}
F(\varepsilon)=\sum\limits_{n=0}^\infty a^\varepsilon_n\cos{n\varepsilon},
\label{eps-series-no-sin}
\end{equation}

Further on, the diurnal Yarkovsky effect must change sign if the asteroid changes its rotation to the opposite,
\begin{equation}
F(\pi-\varepsilon)=-F(\varepsilon)
\label{eps-sign-inversion}
\end{equation}
In this case, the diurnal temperature difference between the leading and the trailing hemispheres has the same value but the different sign, thus pushing the asteroid in the opposite direction with the same force. Substituting $\pi-\varepsilon$ into Eq. (\ref{eps-series-no-sin}), we get
\begin{equation}
F(\pi-\varepsilon)=\sum\limits_{n=0}^\infty a^\varepsilon_n\cos{n(\pi-\varepsilon)}=\sum\limits_{n=0}^\infty a^\varepsilon_n (-1)^n\cos{n\varepsilon},
\end{equation}
As this must be equal to Eq. (\ref{eps-series-no-sin}) with the opposite sign, the coefficients $a^\varepsilon_n$ must vanish for all even $n$, and only odd terms survive in the decomposition.

Ultimately, we get the following expression for $F$ as a function of obliquity:
\begin{equation}
F(\varepsilon)=\sum\limits_{k=0}^\infty a^\varepsilon_{2k+1}\cos{(2k+1)\varepsilon}
\label{eps-series-odd}
\end{equation}
We expect that the terms of the series decay relatively fast, and up to some accuracy the Yarkovsky force can be approximated by its leading Fourier term,
\begin{equation}
F(\varepsilon)\approx a^\varepsilon_1 \cos{\varepsilon}.
\label{eps-diurnal}
\end{equation}

For the seasonal component of the Yarkovsky effect, the same Eq. (\ref{eps-series-no-sin}) holds, as it is derived solely from the symmetry properties of $\varepsilon$. Still, instead of Eq. (\ref{eps-sign-inversion}) we have its opposite, 
\begin{equation}
F(\pi-\varepsilon)=F(\varepsilon).
\label{eps-sign-inversion2}
\end{equation}
It results in cancellation of all odd-order terms in Eq. (\ref{eps-series-no-sin}), and the final expression
\begin{equation}
F(\varepsilon)=\sum\limits_{k=0}^\infty a^\varepsilon_{2k}\cos{2k\varepsilon}
\label{eps-series-even}
\end{equation}
Again, we expect that the high-order Fourier terms will be of lesser importance. Still, we cannot limit ourselves to one leading term, as the seasonal effect must vanish for zero obliquity, thus
\begin{equation}
\sum\limits_{k=0}^\infty a^\varepsilon_{2k}=0.
\end{equation}
We need at least two leading terms to cancel each other, in which case we obtain $a^\varepsilon_{2}=-a^\varepsilon_{0}$, $a^\varepsilon_{4}=a^\varepsilon_{6}=...=0$, and thus
\begin{equation}
F(\varepsilon)\approx a^\varepsilon_0(1- \cos{2\varepsilon})=2 a^\varepsilon_0\sin^2{\varepsilon}.
\label{eps-seasonal}
\end{equation}
As Eqs. (\ref{e-diurnal}) and (\ref{e-seasonal}) were derived for $\varepsilon=0$ and $\pi/2$ respectively, now Eqs. (\ref{eps-diurnal}) and (\ref{eps-seasonal}) imply that they must be multiplied by $\cos\varepsilon$ and $\sin^2\varepsilon$ respectively to approximately account for different values of the latitudes. As a result of all these simplified models, we get the following approximation for the total Yarkovsky force:
\begin{equation}
\Dot{a}=\frac{(1-A)L_{\mathrm{\odot}}}{ Rc\rho\sqrt{M_{\mathrm{\odot}}Ga}}\left(\frac{p_\theta(\theta)\cos\varepsilon}{\pi(1-e^2)}-\frac{p_\theta(\theta_s)\sin^2\varepsilon}{2\pi(1-e^2)}\right).
\label{toy-final}
\end{equation}

The structure of this expression is the following. It represents the Yarkovsky effect as a sum of the diurnal and seasonal components. Each of the components is a product of the terms, depending on the thermal parameter ($\theta$ or $\theta_s$), obliquity ($\varepsilon$), and eccentricity ($e$). Certainly, the both components of the Yarkovsky effect also depend on the asteroid shape. As we assumed a spherical shape of the asteroid throughout our derivation, this dependence is absent in Eq. (\ref{toy-final}), but we expect that different asteroid shapes would result in coefficients other than $\frac{1}{\pi}$ and $\frac{1}{2\pi}$.

\subsection{General formalism}
Recognizing all the imperfections of the simplified theoretical models used to derive Eq. (\ref{toy-final}), we can expect that a more precise expression for the Yarkovsky drift is given by a similar formula but with different functions, which must be determined from a more accurate consideration of the problem:
\begin{align}
\Dot{a}=\frac{(1-A)L_{\mathrm{\odot}}}{Rc\rho\sqrt{M_{\mathrm{\odot}}Ga}}(&p_\theta(\theta)p_s^d(\mathrm{shape})p_\varepsilon^d(\varepsilon)p_e^d(e)-\nonumber\\
-&p_\theta(\theta_s)p_s^s(\mathrm{shape})p_\varepsilon^s(\varepsilon)p_s^d(e)).
\label{general-simple}
\end{align}

The dimensional term in front of the brackets is sufficiently robust to survive through most of the meaningful generalizations of the problem. 

The function $p_\theta$ is in the first approximation the same for the diurnal and seasonal components (with the exception that it depends on two different thermal parameters, $\theta$ and $\theta_s$), but it is possible that in more rigorous consideration $p$ would also split into two different functions, $p^d_\theta(\theta)$ and $p^s_\theta(\theta_s)$. 

For the functions $p^d_s(\mathrm{shape})$ and $p^s_s(\mathrm{shape})$, we expect the dependence on the asteroid shape, but not the asteroid size (except of the small meteoroids with through heat fluxes). This function may strongly depend on the axes ratios of the asteroid, while the other (not too extreme) shape details should produce smaller corrections to this term.

If the asteroid is tilted ($0<\varepsilon<\pi$) and its orbit is eccentric ($e>0$), it is possible that the Yarkovsky effect will depend on the angle $\gamma$ on the asteroid's orbit between the perihelion and the asteroid equinox. This can cause entanglement between these two terms and turn $p_\varepsilon(\varepsilon)p_e(e)$ into one inseparable term $p_{\varepsilon,e}(\varepsilon,e,\gamma)$.

Moreover, the entire concept of factorisation of the Yarkovsky drag into the product of terms depending on individual factors separately, is a mere approximation. For example, the value of the thermal parameter determines how the contribution to the Yarkovsky effect depends on the latitude on the asteroid; on the other hand, non-zero obliquity and eccentricity result in different effective thermal parameters in different points on the orbit. Therefore, a more general expression for the Yarkovsky effect may look like
\begin{align}
\Dot{a}=\frac{(1-A)L_{\mathrm{\odot}}}{Rc\rho\sqrt{M_{\mathrm{\odot}}Ga}}(&p_{\theta,s,\varepsilon,e}^d(\theta,\mathrm{shape},\varepsilon,e,\gamma)-\nonumber\\
-&p_{\theta,s,\varepsilon,e}^s(\theta_s,\mathrm{shape},\varepsilon,e,\gamma)).
\label{yarkovsky-almost-the-most-general}
\end{align}

Furthermore, the mere division of the Yarkovsky effect into the diurnal and seasonal components is but an approximation. In reality, the asteroid's spin rates affect the mean diurnal temperature and thus the seasonal heat flux, influencing in such a way the seasonal Yarkovsky effect. Finally, the most general (and hence the least useful) expression for the Yarkovsky effect may look like
\begin{equation}
\Dot{a}=\frac{(1-A)L_{\mathrm{\odot}}}{Rc\rho\sqrt{M_{\mathrm{\odot}}Ga}}p_{\theta,s,\varepsilon,e}^{d,s}(\theta,\theta_s,\mathrm{shape},\varepsilon,e,\gamma).
\end{equation}

Even this expression still ignores the through heat conduction fluxes (relevant for small meteoroids), the roughness of the surface, the non-constancy of the thermal inertia as a function of temperature and myriads of other delicious and complicated effects that we would rather prefer to leave outside the scope of our consideration for the time being just to avoid being damped in the fine details of the less important factors.

\begin{table}[]
\caption{Initial guesses for the $p$-functions in Eq. (\ref{general-simple})}
\centering
\begin{tabular}{l|l}
Term & Expression \\
\hline
$p_s^d(\mathrm{shape})$ & $\pi^{-1}$\\
$p_s^s(\mathrm{shape})$ & $2\pi^{-1}$\\
$p_\varepsilon^d(\varepsilon)$ & $\cos\varepsilon$\\
$p_\varepsilon^s(\varepsilon)$ & $\sin^2\varepsilon$\\
$p_e^d(e)=p_e^s(e)$ & $(1-e^2)^{-1}$
\end{tabular}
\label{tab:p-different}
\end{table}

For now, we try to stick to the form of Eq. (\ref{general-simple}) as long as possible, until its imperfect factorization starts to substantially compromise its accuracy. We start with the expressions for the individual $p$-terms, as determined in the simplified models of the previous subsections, summarized in Table \ref{tab:p-different}. In the following articles of the series we will improve these expressions one by one, but here we continue with determination the term $p_\theta$, which remained the most important unknown throughout this entire section.

\section{Yarkovsky pressure at the equator}
\label{sec:one_el_equator}
\subsection{Computation of the thermal radiation recoil pressure}
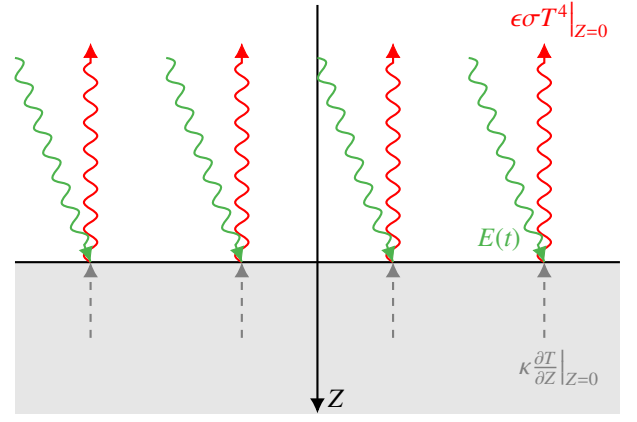
\begin{figure}
\centering
\begin{tikzpicture}
\fill[gray!20!white] (-4,0) rectangle (4,-2);
\draw[thick] (-4,0) -- (4,0);
\draw[thick,-{Latex[length=2mm, width=2mm]}] (0,3.4) -- (0,-2);
\node at (0.25, -1.8) {$Z$};

%IR
\draw[thick,decorate,decoration={coil,aspect=0},red] (-3,0) -- (-3,2.8);
\draw[-{Latex[length=2mm, width=2mm]},red] (-3,2.8) -- (-3,2.9);
\draw[thick,decorate,decoration={coil,aspect=0},red] (-1,0) -- (-1,2.8);
\draw[-{Latex[length=2mm, width=2mm]},red] (-1,2.8) -- (-1,2.9);
\draw[thick,decorate,decoration={coil,aspect=0},red] (1,0) -- (1,2.8);
\draw[-{Latex[length=2mm, width=2mm]},red] (1,2.8) -- (1,2.9);
\draw[thick,decorate,decoration={coil,aspect=0},red] (3,0) -- (3,2.8);
\draw[-{Latex[length=2mm, width=2mm]},red] (3,2.8) -- (3,2.9);
\node[red] at (3.2,3.2) {$\epsilon\sigma T^4\big|_{Z=0}$};

%visible light
\draw[thick,decorate,decoration={coil,aspect=0},teal!60!lime] (-4,2.7) -- (-3,0);
\draw[-{Latex[length=2mm, width=2mm]},teal!60!lime] (-3.01,0.03) -- (-3,0);
\draw[thick,decorate,decoration={coil,aspect=0},teal!60!lime] (-2,2.7) -- (-1,0);
\draw[-{Latex[length=2mm, width=2mm]},teal!60!lime] (-1.01,0.03) -- (-1,0);
\draw[thick,decorate,decoration={coil,aspect=0},teal!60!lime] (0,2.7) -- (1,0);
\draw[-{Latex[length=2mm, width=2mm]},teal!60!lime] (0.99,0.03) -- (1,0);
\draw[thick,decorate,decoration={coil,aspect=0},teal!60!lime] (2,2.7) -- (3,0);
\draw[-{Latex[length=2mm, width=2mm]},teal!60!lime] (2.99,0.03) -- (3,0);
\node[teal!60!lime] at (2.4,0.3) {$E(t)$};

%heat conduction
\draw[-{Latex[length=2mm, width=2mm]},dashed,thick,gray] (-3,-1) -- (-3,0);
\draw[-{Latex[length=2mm, width=2mm]},dashed,thick,gray] (-1,-1) -- (-1,0);
\draw[-{Latex[length=2mm, width=2mm]},dashed,thick,gray] (1,-1) -- (1,0);
\draw[-{Latex[length=2mm, width=2mm]},dashed,thick,gray] (3,-1) -- (3,0);
\node[gray] at (3.2,-1.4) {$\kappa\frac{\partial T}{\partial Z}\big|_{Z=0}$};

\end{tikzpicture}
\caption{Illustration of the geometry of the surface element and the balance of the thermal fluxes on the surface. The gray field is the regolith. The $Z$-axis is directed down. The green waves are the incident solar radiation, the red waves are the emitted infrared flux, and the dashed gray lines are the heat conduction.}
\label{fig:geometry}
\end{figure}

Consider a locally flat element at the equator of a spherical asteroid heated to temperature $T(Z,t)$, which depends on the depth $Z$ below the asteroid surface (see Figure \ref{fig:geometry}) and time $t$. Its surface has the temperature $T|_{Z=0}$ and following the Stefan--Boltzmann law it emits the power $\epsilon\sigma T^4|_{Z=0}$ from a unit surface, where $\sigma$ is the Stefan--Boltzmann constant, and $\epsilon$ is the emissivity, which accounts for the deviation from the ideal black body.

Each emitted photon carries away the momentum $p_1$ equal to its energy $E_1$ divided by the speed of light $c$. Summing up the equation $p_1=E_1/c$ over all photons, and then dividing it by time and surface area, we obtain that emission of radiation in a certain direction causes the recoil force equal to the power divided by $c$. The overall instantaneous recoil pressure experienced by the surface element is therefore
\begin{equation}
P_\mathrm{inst}=\frac{2\epsilon \sigma}{3c} T^4\bigg|_{Z=0}.
\label{P-inst}
\end{equation}
The coefficient $2/3$ accounts for the fact that not all the radiation is emitted perpendicularly to the surface, but also has tangential components that average out producing no net force. The numerical value of the coefficient is obtained by integrating the force over Lambert's emission law, and for different emission laws the coefficient can also be different.

The pressure experienced by the surface element $dS$ creates a force, whose projection on the asteroid orbit equals $F_\mathrm{proj}=P_\mathrm{inst}dS\sin{\phi}$, where $\phi=\omega t$ is the asteroid rotation phase, with $\omega$ being the angular velocity and $t$ being time. (Both $\phi$ and $t$ are measured from the local midday.) Averaging this force over the rotational period of the asteroid, one gets the mean Yarkovsky pressure created by an equatorial surface element
\begin{equation}
P = \frac{\langle F_\mathrm{proj} \rangle}{dS} = \frac{2\epsilon \sigma}{3c} \langle T^4\big|_{Z=0}\sin{\phi}\rangle_\phi.
\label{P_dimensional}
\end{equation}
For the computation of this Yarkovsky pressure, surface temperature $T$ needs to be evaluated, and the thermal model of the asteroid soil needs to be constructed.

\subsection{Heat conduction equation}
We use the thermal model described by \cite{lagerros1996}.
The one-dimensional heat diffusion equation below the flat patch of the asteroid surface has the form
\begin{equation}
\label{conductiv}
\frac{\partial T}{\partial t} = \frac{\kappa}{C\rho} \frac{\partial^2 T}{\partial Z^2},
\end{equation}
where $\kappa$ is the thermal conductivity, $C$ is the heat capacity, and $\rho$ is the bulk density of the asteroid surface layer. Throughout this paper we assume that $\kappa$, $C$, and $\rho$ are constants, independent on either $T$ or $Z$.

The surface boundary condition for this equation is (see Figure \ref{fig:geometry})
\begin{equation}
\label{boundary}
\kappa \frac{\partial T}{\partial Z}\bigg|_{Z=0} =  \epsilon\sigma\ T^4\bigg|_{Z=0} - E,
\end{equation}
with $E$ being the impinging solar power absorbed by the unit surface:
\begin{equation}
E=\frac{(1-A)L_\odot}{4\pi r^2}\alpha
\label{E_of_alpha}
\end{equation}
Here $A$ is the bolometric Bond albedo of the asteroid, $L_\odot=3.828\cdot 10^{26}$ W is the solar luminosity, $r$ is the asteroid's heliocentric distance, and $\alpha$ is a non-dimensional geometric factor
\begin{equation}
\alpha(\phi)=\cos{\phi}\cdot H(\cos{\phi}),
\label{alpha}
\end{equation}
with $H$ denoting the Heaviside step-function.

The other boundary condition is set at a great depth
\begin{equation}
\label{boundary2}
\kappa \frac{\partial T}{\partial Z}\bigg|_{Z \rightarrow \infty} = 0.
\end{equation}
%This condition is true only if the surface of the asteroid does not curve substantially on the distance scale of a few thermal wavelengths.
This condition is true only in the case of a flat plane approximation when the surface of the asteroid does not curve substantially on the distance scale of a few thermal wavelengths.

Assuming that the asteroid's day is much shorter than the asteroid's year, we conclude that the temperature is periodic with the period of the asteroid's rotation $2\pi/\omega$:
\begin{equation}
\label{boundary3}
T\big|_{t=2\pi/\omega} = T\big|_{t=0}.
\end{equation}
For the heat conduction equation, this periodicity condition substitutes the initial condition.

The characteristic scales of the temperature and the distance are given by
%\begin{equation}
%l=\sqrt{\frac{\kappa}{C\rho\omega}},\,\,T_\mathrm{SS}=\sqrt[4]{\frac{\left(1 - A\right)L_\odot}{\epsilon \sigma\cdot 4\pi r^2}},
%\label{skin_depth_subsol_tempr}
%\end{equation}

\begin{equation}
T_\mathrm{SS}=\sqrt[4]{\frac{\left(1 - A\right)L_\odot}{\epsilon \sigma\cdot 4\pi r^2}},
\label{subsol_tempr}
\end{equation}

\begin{equation}
l=\sqrt{\frac{\kappa}{C\rho\omega}},
\label{skin_depth}
\end{equation}
respectively, and the physical meaning of $T_\mathrm{SS}$ is the temperature of the sub-solar point, whereas $l$ is of the order of the thermal wavelength\footnote{Traditionally, in the theory of the Yarkovsky effect, it is accepted to define bodies as thermally ``small'' or ``large'' in terms of the ratio of the object radius $R$ to the characteristic penetration depth $l$ of temperature changes \citep{peterson1976a, burns_etal1979, vokrouhlicky1998a}. Here, we only consider thermally ``large'' objects, i.e. $R/l \rightarrow \infty$.}.

It is convenient to non-dimensionalize the equations, using the following non-dimensional variables:
\begin{equation}
\phi = \omega t,\,\,
z = Z/l,\,\,
\tau = T/T_\mathrm{SS}.
\label{tau}
\end{equation}
Then Eqs. (\ref{conductiv}), (\ref{boundary}), (\ref{boundary2}), and (\ref{boundary3}) transform into the following non-dimensional form:
\begin{equation}
\label{pde}
\frac{\partial \tau}{\partial \phi} = \frac{\partial^2 \tau}{\partial z^2},
\end{equation}
\begin{equation}
\label{boundary_condition}
\theta \frac{\partial \tau}{\partial z}\bigg|_{z=0} = \tau^4\bigg|_{z=0}-\alpha(\phi),
\end{equation}
\begin{equation}
\label{boundary_condition2}
\frac{\partial \tau}{\partial z}\bigg|_{z \rightarrow \infty} = 0,
\end{equation}
\begin{equation}
\label{boundary_condition3}
\tau\big|_{\phi=2\pi} = \tau\big|_{\phi=0}.
\end{equation}
The thermal parameter $\theta$, which enters these equations, is defined as
\begin{equation}
\theta=\frac{\sqrt{C\rho\kappa\omega}}{\epsilon\sigma T_\mathrm{SS}^3}=\frac{(C\rho\kappa\omega)^{1/2}}{(\epsilon\sigma)^{1/4}}\left(\frac{4\pi r^2}{(1-A)L_\odot}\right)^{3/4}
\label{theta-def}
\end{equation}
This $\theta$ is the single most important parameter studied in this article, hence ``thermal dependence'' in its name. It incorporates all the information about the thermal properties of the soil, the distance to the Sun, the rotation rate, the albedo, and the emissivity. Each value of $\theta$ produces a family of self-similar solutions to the thermal problem, and there is no need to model different values of all parameters separately, as long as the dimensionless number $\theta$ remains constant.

A high value of $\theta$ implies a high thermal inertia and a small temperature change over the rotation period. On the contrary, low $\theta$ implies an almost instantaneous adjustment of the surface temperature to changing solar radiation, and thus an extreme daily temperature change. The typical values of $\theta$ for different asteroids will be discussed in Section \ref{sec:discussion}.

With all the variables already being non-dimensionalized, it is convenient to non-dimensionalize the Yarkovsky pressure $P$ from Eq. (\ref{P_dimensional}) as well. The characteristic value of the pressure is given by the energy flux absorbed by the sub-solar point, divided by the speed of light,
\begin{equation}
P_0=\frac{(1-A)L_\odot}{4\pi r^2 c}.
\label{P0-def}
\end{equation}
%\begin{equation}
%F_\mathrm{Y}=\frac{(1-A)L_\odot S}{4\pi r^2 c}p.
%\end{equation}
Then the non-dimensional pressure is defined as
\begin{equation}
p_\theta=\frac{P}{P_0} = \frac{2}{3} \langle \tau^4\big|_{z=0}\sin{\phi}\rangle_\phi.
\label{p_non-dimensional}
\end{equation}
Comparing this definition with Eq. (\ref{p_theta_def1}), we see that this is the same $p_\theta$, which was earlier used in the derivation of Eqs. (\ref{toy-final}) and (\ref{general-simple}).

A straightforward albeit not always practical approach to this heat conduction problem is to solve the set of equations (\ref{pde})--(\ref{boundary_condition3}) numerically. For this purpose, we apply a finite-difference scheme based on the Crank--Nicolson method to a one-dimensional slab of asteroid material reaching the depth of several thermal wavelengths beneath the surface. The boundary condition is tackled by one Newtonian iteration at each time step. Uniform temperature distribution is taken for the initial condition, but several dozen revolution cycles of the asteroid are given for equilibrization, after which the mean Yarkovsky force is computed over the last revolution cycle. The time step, the spacial step, the depth of the slab, and the number of equilibrization cycles are selected as a result of numerical tests to provide the relative error of the resulting Yarkovsky force less than 0.001 in the range of thermal parameters $0.001<\theta<1000$.

Still, it is possible to construct approximate analytical solutions for Eqs. (\ref{pde})-(\ref{boundary_condition3}) for the cases $\theta \gg 1$ and $\theta \ll 1$, and compute $p_\theta$ in these cases. These analytical approaches provide important limiting cases for the numerical modeling. We present them in the following subsections.

\subsection{Case of large thermal parameter ($\theta \gg 1$)}
In this case, the daily change of the temperature is small.
The mean 4$^\mathrm{th}$-power dimensionless temperature $\tau_0^4=\langle\tau^4\rangle_\phi$ can be estimated by averaging Eq. (\ref{boundary_condition}) over $\phi$. As the net daily heat flux into the depth of the asteroid is zero (Eq. (\ref{boundary_condition2})), and as there is no net change of the thermal energy accumulated in the soil (Eq. (\ref{boundary_condition3})), the net heat flux through the asteroid surface should also vanish. Then from Eq. (\ref{boundary_condition}) we get
\begin{equation}
\tau_0^4=\frac{1}{2\pi}\int\limits_0^{2\pi}\alpha(\phi) d\phi=\frac{1}{2\pi}\int\limits_{-\pi/2}^{\pi/2}\cos{\phi}\,d\phi=\frac{1}{\pi}
\label{tau_0}
\end{equation}
Next consider a small deviation of the temperature from this mean value, $\tau=\tau_0+\Delta\tau$. Substituting this expression into Eq. (\ref{boundary_condition}) and neglecting the higher-order terms in terms of $\Delta\tau$, we get a linearized form of the boundary condition:
\begin{equation}
\theta\frac{\partial\Delta\tau}{\partial z}\bigg|_{z=0}= \tau_0^4+4\tau_0^3\Delta\tau\bigg|_{z=0}-\alpha(\phi).
\label{boundary_condition_linear}
\end{equation}
As the non-linearity $\tau^4$ was the major complication of the original set of heat conduction equations, having eliminated this non-linearity we obtain a much simpler problem that can be solved by standard methods of the Fourier analysis. The non-dimensional temperature perturbation can be expressed as the sum of all the possible decaying temperature waves propagating into the depth of the asteroid that have $2\pi$ as one of their periods:
\begin{align}
\Delta\tau=\sum\limits_{n=0}^{\infty}&\exp\left(-\sqrt{\frac{n}{2}}z\right)\times\nonumber\\ &\times\left(a_n\cos\left(n\phi-\sqrt{\frac{n}{2}}z\right)+b_n\sin\left(n\phi-\sqrt{\frac{n}{2}}z\right)\right).
\label{d_tau_Fourier}
\end{align}
By construction, this expression satisfies the heat conduction equation Eq. (\ref{pde}), its in-depth boundary condition Eq. (\ref{boundary_condition2}), and its periodic initial condition Eq. (\ref{boundary_condition3}). Therefore, the surface boundary condition Eq. (\ref{boundary_condition_linear}) remains the only equation that needs to be satisfied to determine the Fourier coefficients $a_n$ and $b_n$:
\begin{align}
\theta\sum\limits_{n=0}^{\infty} \sqrt{\frac{n}{2}}(a_n\sin{n\phi} -b_n\cos{n\phi}-a_n\cos{n\phi}-b_n\sin{n\phi})=\nonumber\\
=\tau_0^4+4\tau_0^3\sum\limits_{n=0}^{\infty} (a_n\cos{n\phi}+b_n\sin{n\phi})-\alpha(\phi).
\label{lin-bound-cond}
\end{align}
We also decompose $\alpha$ into the Fourier series,
\begin{equation}
\alpha(\phi)=\sum\limits_{n=0}^{\infty} A_n \cos{n\phi},
\label{Fourier_An}
\end{equation}
with $A_0=\frac{1}{\pi}$,
$A_1=\frac{1}{2}$, all other odd-order coefficients being zeros ($A_{2k+1}=0$, $k\ge 1$), and all other even-order coefficients given as
\begin{equation}
A_{2k} = \frac{2(-1)^{k+1}}{\pi(4k^2-1)},\,\,\,k\ge 1.
\end{equation}
Balancing terms in front of the same harmonics in Eq. (\ref{lin-bound-cond}) results for each $n\ge 1$ into the set of two linear equations with the two unknowns, $a_n$ and $b_n$:
\begin{equation}
\begin{cases}
\displaystyle\theta\sqrt{\frac{n}{2}} a_n-\theta\sqrt{\frac{n}{2}}b_n=4\tau_0^3 b_n, \\
\\
\displaystyle -\theta\sqrt{\frac{n}{2}}b_n-\theta\sqrt{\frac{n}{2}}a_n=4\tau_0^3 a_n-A_n.
\end{cases}
\label{set_of_eqs_for_b1}
\end{equation}
Solving this set of equations, we obtain
\begin{equation}
a_n=\frac{A_n\left(\theta+4\tau_0^3\sqrt{\frac{2}{n}}\right)\sqrt{\frac{2}{n}}}{\theta^2+\left(\theta+4\tau_0^3\sqrt{\frac{2}{n}}\right)^2},\,\,\,\,
b_n=\frac{A_n\theta\sqrt{\frac{2}{n}}}{\theta^2+\left(\theta+4\tau_0^3\sqrt{\frac{2}{n}}\right)^2}.
\label{b_1}
\end{equation}

Substituting $\tau=\tau_0+\Delta\tau$ into  Eq. (\ref{p_non-dimensional}) and using the Fourier decomposition Eq. (\ref{d_tau_Fourier}), we see that all the oscillating trigonometric functions vanish after averaging, and the only non-vanishing term in Eq. (\ref{p_non-dimensional}) is the one with $b_1$:
\begin{align}
p_\theta&=\frac{2}{3}\langle(\tau_0^4+4\tau_0^3\Delta\tau)\big|_{z=0}\sin{\phi}\rangle_\phi=\nonumber\\
&=\frac{8\tau_0^3}{3}\left\langle\sum\limits_{n=1}^{\infty} (a_n\cos{n\phi}+b_n\sin{n\phi})\cdot\sin{\phi}\right\rangle_\phi=\nonumber\\
&=\frac{8\tau_0^3}{3}\left\langle b_1\sin{^2\phi}\right\rangle_\phi=\frac{4\tau_0^3 b_1}{3}
\label{p_computation}
\end{align}
Substituting $\tau_0$ from Eq. (\ref{tau_0}) and $b_1$ from Eq. (\ref{b_1}), we get the final expression for the Yarkovsky pressure:
\begin{align}
p_\theta=\frac{1}{6}&\cdot\frac{\frac{\pi^{3/4}}{4\sqrt{2}}\theta}{1+2\cdot\frac{\pi^{3/4}}{4\sqrt{2}}\theta+2\cdot\left(\frac{\pi^{3/4}}{4\sqrt{2}}\theta\right)^2}=\nonumber\\
&=\frac{\theta}{\frac{24\sqrt{2}}{\pi^{3/4}}+12\cdot\theta+\frac{3}{\sqrt{2}}\pi^{3/4}\cdot\theta^2}
\label{p_big_theta}
\end{align}

\subsection{Analysis of the expression for large thermal parameter}

Substituting Eq. (\ref{p_big_theta}) into Eq. (\ref{p_theta_def1}) we obtain the same expression for Yarkovsky force as in the paper of \cite{peterson1976a} (see his Eq. (17b) for $F_y$, where, to a first approximation, parameter $P$ can be found as $P=\theta \cdot \pi^{3/4}/\sqrt{2}$). In that paper author solves the heat equation in the same way as we have done in previous subsection, just only without non-dimensionalized variables, also hinting that such solution will break down in case of small $\theta$. Our Eq. (\ref{p_big_theta}) also agrees with the solution for the diurnal Yarkovsky effect derived by \cite{vokrouhlicky1998a} using spherical Bessel functions, if the limiting case $R\gg l$ is considered.

\begin{figure}
\begin{center}
\includegraphics[width=.49\textwidth]{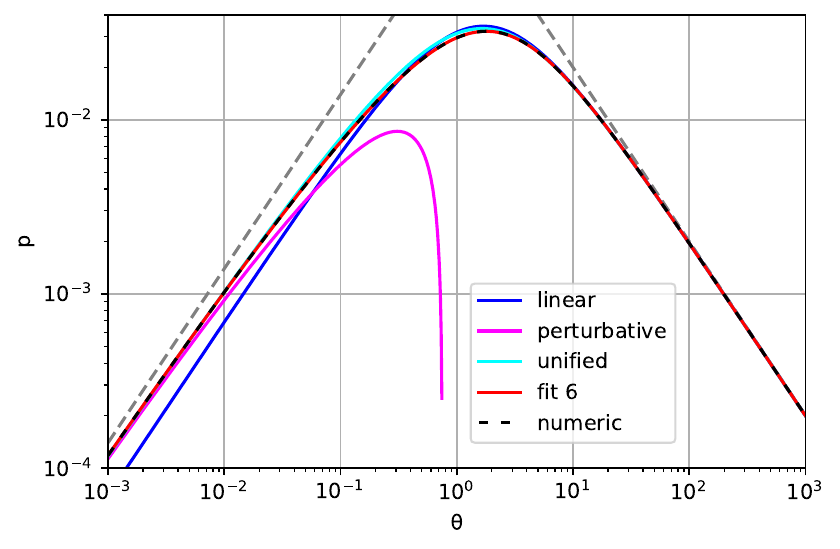}
\caption{The non-dimensional Yarkovsky pressure $p$ as a function of the thermal parameter $\theta$.}
\label{fig:p(theta)}
\end{center}
\end{figure}

In Figure \ref{fig:p(theta)} we see that Eq. (\ref{p_big_theta}) indeed agrees very well with the results of numerical simulations for $\theta\gtrsim 10$. For large $\theta$ the denominator is dominated by its first term, and the equation reduces to

\begin{equation}
%p_\theta\approx\frac{3\pi^{3/4}}{\sqrt{2}\theta}\approx 3.7599\cdot\theta^{-1}.
p_\theta\approx\frac{\sqrt{2}}{3\pi^{3/4}\theta}\approx 0.19977\cdot\theta^{-1}.
\label{p_very_big_theta}
\end{equation}
This expression is shown with a dashed gray line in the right-hand side of Figure \ref{fig:p(theta)}. In logarithmic coordinates, this power-law prescribes a straight line that asymptotically approaches the exact expression at large $\theta$. It is almost indistinguishable from the numerical results for $\theta\gtrsim 100$, although deviates from it for smaller $\theta$.

\cite{radzievskii1952} and \cite{peterson1976a} clearly understood the limitation of the linear solution and proposed to use it to evaluate the Yarkovsky effect only for objects with large $\theta$. Unfortunately, the later authors forgot to mention this assumption clearly and even used the linearized model outside the range of its applicability. Some authors \citep{peterson1976b,sekiya_etal2012} have attempted to improve the solution by accounting for higher-order terms from the Taylor expansion, finally concluding that they do not play a significant role in refining the magnitude of the Yarkovsky effect, and that the linear approximation can still be used only for $\theta \gtrsim 10$.

Although this section deals with the diurnal effect, it is important to note that, for the seasonal effect, the linear theory is equally inadequate in the region of small $\theta$. \cite{rubincam1995} estimated the magnitude of the diurnal temperature fluctuation on the surface of an asteroid, calculated using a linear approach, then compared his results with the results of numerical modeling presented in \cite{spencer1989} and arrived at the conclusion that the linear model overestimates the diurnal Yarkovsky effect by approximately 25\% in the case of $\theta$ = 1, and the same magnitude of error should be expected for the seasonal effect. As a result, Rubincam added a factor of 0.75 to the formulas for the seasonal rate of orbital evolution (\citeauthor{rubincam1995} \citeyear{rubincam1995}, \citeyear{rubincam1998}), underestimating the result for all range of $\theta$.
\cite{vokrouhlicky_farinella1998}, like Rubincam, found discrepancies of about 20\% between linear theory and numerical simulations for the seasonal effect, but concluded that such a discrepancy becomes critical only in the case of significantly eccentric orbits, and that the linear approach is still able to predict well long-term changes in mean elements averaged over the orbit.

So one should keep in mind, that Eq. (\ref{p_big_theta}), obtained under the assumption of small fluctuations of surface temperature relative to its average value, is asymptotically correct only for $\theta\gg 1$, and using it for small thermal parameters (as it almost always happens in literature) is unjustified.

\subsection{Case of small thermal parameter ($\theta \ll 1$)}
In this subsection, we develop the perturbation theory for small $\theta$, following the approach proposed by \citep{golubov2016}. In the 0$^\mathrm{th}$-order approximation, we neglect the $\theta$ term in Eq. (\ref{boundary_condition}) altogether, obtaining the 0$^\mathrm{th}$ approximation for the surface temperature:
\begin{equation}
\tau_0\big|_{z=0}=\sqrt[4]{\alpha(\phi)}.
\label{tau_0_surf}
\end{equation}
(One must keep in mind that this $\tau_0$ is completely different from $\tau_0$ in Eq. (\ref{tau_0}) -- these are 0$^\mathrm{th}$ approximations in two different approaches.)
The general solution at arbitrary depth is again expressed by the Fourier series,
\begin{align}
\tau_0=\sum\limits_{n=0}^{\infty}& \exp\left(-\sqrt{\frac{n}{2}}z\right)\times\nonumber\\
&\times\left(a_n\cos\left(n\phi-\sqrt{\frac{n}{2}}z\right)+b_n\sin\left(n\phi-\sqrt{\frac{n}{2}}z\right)\right)
\label{tau_0_depth}
\end{align}
By equating Eq. (\ref{tau_0_depth}) at $z=0$ to Eq. (\ref{tau_0_surf}), we deduce that $a_n$ and $b_n$ are simply the coefficients of the Fourier decomposition of $\sqrt[4]{\alpha(\phi)}$,
\begin{equation}
a_0=\frac{1}{2\pi}\int\limits_0^{2\pi}\sqrt[4]{\alpha(\phi)}d\phi
\label{a_0}
\end{equation}
\begin{equation}
a_n=\frac{1}{\pi}\int\limits_0^{2\pi}\sqrt[4]{\alpha(\phi)}\cos{n\phi}d\phi\,\,\,(n\neq 0)
\label{a_n_2}
\end{equation}
\begin{equation}
b_n=\frac{1}{\pi}\int\limits_0^{2\pi}\sqrt[4]{\alpha(\phi)}\sin{n\phi}d\phi = 0
\end{equation}
Due to the absence of $b_1$ term, the Yarkovsky force is absent in this 0$^\mathrm{th}$-order approximation, and the next iteration needs to be carried out. So we substitute the 0$^\mathrm{th}$-order approximation $\tau_0$ into the left-hand side of Eq. (\ref{boundary_condition}), and read the next iteration $\tau_1$ from the right-hand side:
\begin{equation}
\tau_1^4\bigg|_{z=0}=\theta\frac{\partial\tau_0}{\partial z}\bigg|_{z=0}+\alpha(\phi)
\label{bondary_condition_1st_iteration}
\end{equation}
This surface value of $\tau_1$ can be used to obtain the 1$^\mathrm{st}$ approximation for the dimensionless Yarkovsky pressure:
\begin{equation}
p_\theta=\frac{2}{3}\left\langle\left(\alpha(\phi)+\theta\frac{\partial\tau_0}{\partial z}\bigg|_{z=0}\right)\sin{\phi}\right\rangle_\phi
\end{equation}
The even function $\alpha(\phi)$ multiplied by the odd function $\sin\phi$ cancels after averaging, whereas $\tau_0$ is substituted from Eq. (\ref{tau_0_depth}):
\begin{equation}
p_\theta=\frac{2\theta}{3}\left\langle\sum\limits_{n=0}^{\infty} \sqrt{\frac{n}{2}}(a_n\sin{n\phi} -a_n\cos{n\phi})\sin{\phi}\right\rangle_\phi = \frac{\theta}{3\sqrt{2}}a_1,
\label{p_small_theta_intermediate}
\end{equation}
as only the $n=1$ term survives the averaging. Substituting $a_1$ from Eq. (\ref{a_n_2}) and $\alpha$ from Eq. (\ref{alpha}), we get
\begin{equation}
p_\theta=-\frac{4\sqrt{2}\Gamma\left(\frac{1}{8}\right)}{45\sqrt{\pi}\Gamma\left(-\frac{3}{8}\right)}\theta \approx 0.13968\cdot\theta,
\label{p_small_theta}
\end{equation}
where $\Gamma$ if the gamma-function.

\subsection{Analysis of the expression for small thermal parameter}
It is instructive to compare this expression with the $\theta\rightarrow 0$ limit of Eq. (\ref{p_big_theta}), which is
\begin{equation}
p_\theta\approx\frac{\pi^{3/4}}{24\sqrt{2}}\theta \approx 0.06952\cdot\theta
\label{p_small_theta_wrong}
\end{equation}
We see that Eqs. (\ref{p_small_theta}) and (\ref{p_small_theta_wrong}) differ by the factor of approximately 2. Therefore, using the traditional expression Eq. (\ref{p_big_theta}) outside its applicability range unacceptably underestimates the Yarkovsky effect.
The same can be seen in the left-hand side of Figure \ref{fig:p(theta)}. The ``classical'' theory expressed by Eq. (\ref{p_big_theta}) significantly deviates from the numerical simulations.

Although this approach to the case of $\theta\ll 1$ seems simple and straightforward, it has an important caveat. The issue is that the zeroth iteration for the surface temperature given by Eq. (\ref{tau_0_surf}) has infinite derivatives at sunset and sunrise. It results in infinite derivatives $\partial\tau_0/\partial z|_{z=0}$. After substitution into Eq. (\ref{bondary_condition_1st_iteration}), it produces infinite estimates for $\tau_1^4|_{z=0}$: however small the $\theta$, the product of small $\theta$ by infinite $\partial\tau_0/\partial z|_{z=0}$ is still infinite. For small $\theta$ the discontinuity of $\tau_1^4|_{z=0}$ shrinks to ever smaller temporal regions around the sunset and sunrise, but never vanishes. Even worse, close to the sunset and sunrise $\tau_1^4|_{z=0}$ acquires not only positive, but also negative values, implying complex values of $\tau_1|_{z=0}$. In Eq. (\ref{p_small_theta_intermediate}), the infinities miraculously cancel out, but their mere presence is a sufficient source of doubt in the entire method. We postpone the rigorous proof of this method for the future, and for now just apply it heuristically as it is, while the good agreement between the analytic expression and the numeric simulation serves for us as the main proof of the analytic method's solidity. Still, to make the agreement even better, we go on and construct the next iteration within the same formalism.

\subsection{Case of small thermal parameter. Second iteration}
To conduct this next iteration, we define the surface temperature $\tau_1|_{z=0}$ from Eq. (\ref{bondary_condition_1st_iteration}), with $\tau_0$ being substituted from Eq. (\ref{tau_0_depth}):
\begin{equation}
\tau_1^4\bigg|_{z=0}=\theta\sum\limits_{n=1}^{\infty} \sqrt{\frac{n}{2}}a_n(\sin{n\phi} -\cos{n\phi})+\alpha(\phi)
\end{equation}
At night, for $\pi/2<\phi<3\pi/2$, the insolation is zero, $\alpha=0$, and we get
\begin{equation}
\tau_{1,\mathrm{night}}\bigg|_{z=0}=\theta^{1/4}\left(\sum\limits_{n=1}^{\infty} \sqrt{\frac{n}{2}}a_n(\sin{n\phi} -\cos{n\phi})\right)^{1/4}
\label{night}
\end{equation}
At day, for $-\pi/2<\phi<\pi/2$, the insolation $\alpha$ is non-zero, and most time it is of the order of unity. As $\theta$ is assumed to be small, we can Taylor-decompose this expression into
\begin{equation}
\tau_{1,\mathrm{day}}\bigg|_{z=0}=\sqrt[4]{\alpha(\phi)}+O(\theta),
\label{day}
\end{equation}
where $O(\theta)$ stands for all the terms of the order of $\theta$.

Equations (\ref{night}) and (\ref{day}) give the surface temperature at any point in time. But to construct the next iteration $\tau_2$ we need to know the temperature $\tau_1$ at depth. As it is a solution of the heat conduction equation, it can only be the sum of temperature waves:
\begin{align}
\tau_1=\sum\limits_{n=0}^\infty&\exp\left(-\sqrt{\frac{n}{2}}\right)\times\nonumber\\
&\times\left(\tilde{a}_n\cos\left(n\phi-\sqrt{\frac{n}{2}}\right)+\tilde{b}_n\sin\left(n\phi-\sqrt{\frac{n}{2}}\right)\right)
\label{tau1_general}
\end{align}
On the surface, the temperature from Eq. (\ref{tau1_general}) should be equal to Eqs. (\ref{night}) and (\ref{day}) at night and day correspondingly. 
This condition allows us to determine the coefficients of the Fourier decomposition. In particular, the coefficients $\tilde{a}_1$ and $\tilde{b}_1$ that we will need further are as follows:
\begin{align}
&\tilde{a}_1=\frac{1}{\pi}\int\limits_0^{2\pi}\tau_1\bigg|_{z=0}\cos\phi\,d\phi=\frac{1}{\pi}\int\limits_{-\pi/2}^{\pi/2}\sqrt[4]{\alpha(\phi)}\cos\phi\,d\phi+O(\theta)+\nonumber\\
&+\frac{1}{\pi}\int\limits_{\pi/2}^{3\pi/2}\theta^{1/4}\left(\sum\limits_{n=1}^{\infty} \sqrt{\frac{n}{2}}a_n(\sin{n\phi} -\cos{n\phi})\right)^{1/4}\cos\phi\,d\phi=\nonumber\\
&=\frac{\theta^{1/4}}{\pi}\int\limits_{\pi/2}^{3\pi/2}\left(\sum\limits_{n=1}^{\infty} \sqrt{\frac{n}{2}}a_n(\sin{n\phi} -\cos{n\phi})\right)^{1/4}\cos\phi\,d\phi+\nonumber\\
&\,\,\,\,\,\,\,\,\,\,\,\,\,\,\,\,\,\,\,\,\,\,\,\,\,\,\,\,\,\,\,\,\,\,\,\,\,\,\,\,\,\,\,\,\,\,\,\,\,\,\,\,\,\,\,\,\,\,\,\,\,\,\,\,\,\,\,\,\,\,\,\,\,\,\,\,\,\,\,\,\,\,\,\,\,\,\,\,\,\,\,\,\,\,\,\,\,\,\,\,\,\,\,\,\,+a_1+O(\theta),
\label{a_tilde_1}
\end{align}
\begin{align}
&\tilde{b}_1= \frac{1}{\pi}\int\limits_0^{2\pi}\tau_1\bigg|_{z=0}\sin\phi\,d\phi=\frac{1}{\pi}\int\limits_{-\pi/2}^{\pi/2}\sqrt[4]{\alpha(\phi)}\sin\phi\,d\phi+O(\theta)+\nonumber\\
&+\frac{1}{\pi}\int\limits_{\pi/2}^{3\pi/2}\theta^{1/4}\left(\sum\limits_{n=1}^{\infty} \sqrt{\frac{n}{2}}a_n(\sin{n\phi} -\cos{n\phi})\right)^{1/4}\sin\phi\,d\phi=\nonumber\\
&=\frac{\theta^{1/4}}{\pi}\int\limits_{\pi/2}^{3\pi/2}\left(\sum\limits_{n=1}^{\infty} \sqrt{\frac{n}{2}}a_n(\sin{n\phi} -\cos{n\phi})\right)^{1/4}\sin\phi\,d\phi+O(\theta).
\label{b_tilde_1}
\end{align}
Here we made use of the fact that the coefficients $a_n$ defined by Eqs. (\ref{a_0}) and (\ref{a_n_2}) give the Fourier decomposition of $\sqrt[4]{\alpha(\phi)}$.

We substitute $\tau_1$ defined by Eq. (\ref{tau1_general}) into the right-hand side of the boundary condition Eq. (\ref{boundary_condition}), and read the next iteration for the surface temperature $\tau_2$ from the left-hand side of the equation:
\begin{equation}
\tau_2^4\bigg|_{z=0}=\theta\frac{\partial\tau_1}{\partial z}\bigg|_{z=0}+\alpha(\phi)    
\end{equation}
Then we substitute this $\tau_2$ into Eq. (\ref{p_non-dimensional}) and compute the next iteration for the Yarkovsky pressure $p_\theta$:
\begin{equation}
p_\theta=\frac{2}{3}\langle\tau_2^4\big|_{z=0}\sin{\phi}\rangle_\phi
\end{equation}
After averaging with the factor $\sin{\phi}$, the even term $\alpha(\phi)$ vanishes. The Fourier terms of $\tau_1$ with $\sin{n\phi}$ and $\cos{n\phi}$ that are orthogonal to $\sin{\phi}$ also vanish, and the only two non-vanishing terms are the ones with $\tilde{a}_1\sin{\phi}$ and $\tilde{b}_1\sin{\phi}$:
\begin{align}
p_\theta&=\frac{2}{3}\left\langle\theta\frac{\partial\tau_1}{\partial z}\bigg|_{z=0}\sin{\phi}\right\rangle_\phi=\nonumber\\
&=\frac{2}{3}\left\langle\theta \frac{1}{\sqrt{2}}(\tilde{a}_1\sin{\phi}-\tilde{b}_1\sin{\phi})\sin{\phi}\right\rangle_\phi=\frac{\theta}{3\sqrt{2}}(\tilde{a}_1-\tilde{b}_1)
\end{align}
Substituting $\tilde{a}_1$ and $\tilde{b}_1$ from Eqs. (\ref{a_tilde_1}) and (\ref{b_tilde_1}), and neglecting $O(\theta)$ terms in comparison with $O(\theta^{1/4})$ terms for small $\theta$, we obtain:
\begin{align}
p_\theta=\frac{a_1 \theta}{3\sqrt{2}}\bigg(1+\frac{\theta^{1/4}}{\pi a_1}\int\limits_{\pi/2}^{3\pi/2}\left(\sum\limits_{n=1}^{\infty} \sqrt{\frac{n}{2}}a_n(\sin{n\phi} -\cos{n\phi})\right)^{1/4}\cdot\nonumber\\
\cdot(\cos\phi-\sin\phi)\,d\phi\bigg)
\label{p_not_so_small_theta_intermediate}
\end{align}
Numerically computing the integral of the sum, we get
\begin{align}
p_\theta&=-\frac{4\sqrt{2}\Gamma\left(\frac{1}{8}\right)}{45\sqrt{\pi}\Gamma\left(-\frac{3}{8}\right)}\cdot\theta\cdot(1-0.88935\cdot\theta^{1/4}) \approx\nonumber\\
&\approx 0.13968\cdot\theta\cdot(1-0.88935\cdot\theta^{1/4}).
\label{p_not_so_small_theta}
\end{align}

For very small $\theta$ this expression reduces to Eq. (\ref{p_small_theta}). Nevertheless, as the relative difference between Eqs. (\ref{p_small_theta}) and (\ref{p_not_so_small_theta}) is proportional to $\theta^{1/4}$, it remains noticeable even for relatively small $\theta$. If we trace back the origin of this $\theta^{1/4}$ factor through the equations, we find that ultimately it comes from the night temperature on the asteroid. Even for very small $\theta$, the night temperature cannot reduce too fast due to the power of 4 on the Stefan--Boltzmann law, and this non-zero night temperature proportional to $\theta^{1/4}$ produces the major correction to the Yarkovsky effect, whereas the correction arising from the day temperature is much smaller and proportional to $\theta$.

This qualitative picture is illustrated graphically in Figure \ref{fig:p(theta)}. The deviation of the numerical solution from its left-hand side asymptotic Eq. (\ref{p_small_theta}) remains substantial much longer than its deviation from the right-hand side asymptotic Eq. (\ref{p_very_big_theta}), as the latter deviation is defined by $\theta$, whereas the former by its power 1/4. This different rate of convergence gives the plot its asymmetric lopsided appearance. The second iteration provided by Eq. (\ref{p_not_so_small_theta}) describes the major part of the deviation of the numerical solution from the asymptotic. Naturally, this formula works well only for $\theta\ll 1$, whereas for $\theta\sim 1$ it substantially misses the numerical solution, and for $\theta>1.6$ it gives negative $p_\theta$. (This change of sign of $p_\theta$ is seen as a near-vertical ``tail'' on the curve.)

\subsection{Unified expression for the Yarkovsky force}
We have Eq. (\ref{p_big_theta}) that describes $p_\theta$ at $\theta\gg 1$ and Eq. (\ref{p_not_so_small_theta}) that describes it at $\theta\ll 1$. Now we want to sew these two equations into one Frankenstein expression valid for both small and big $\theta$.
The following analytical equation does just that:
\begin{align}
p_\theta=\frac{\theta}{-\frac{45\sqrt{\pi}\Gamma\left(-\frac{3}{8}\right)}{4\sqrt{2}\Gamma\left(\frac{1}{8}\right)}+\frac{675\Gamma^2\left(-\frac{3}{8}\right)}{16\sqrt{2}\Gamma^2\left(\frac{1}{8}\right)}\theta^{1/4}+12\theta+\frac{3\pi^{3/4}}{\sqrt{2}}\theta^2}
\label{p_unified_analytic}
\end{align}
Indeed, for $\theta\gg 1$, the denominator of the equation is dominated by the first two terms, which makes it equivalent to Eq. (\ref{p_not_so_small_theta}). On the other hand, for $\theta\ll 1$, the last two terms play the major role in the denominator, and Eq. (\ref{p_unified_analytic}) reduces to an asymptotic equivalent of Eq. (\ref{p_not_so_small_theta}).

\begin{figure}
\centering
\includegraphics[width=\linewidth]{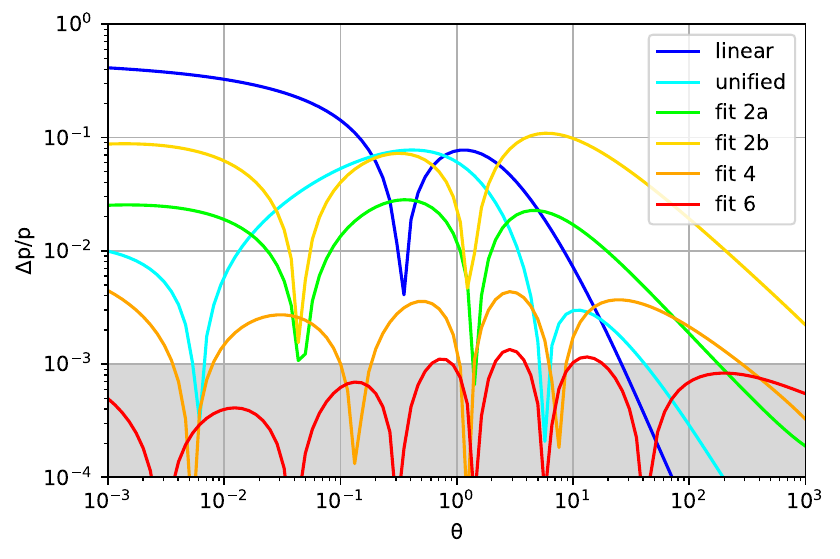}
\caption{Error of different analytical approximations $p_\mathrm{a}(\theta)$ with respect to the numerical solution $p_\mathrm{n}(\theta)$. The vertical axis shows the absolute value of the relative error, $|p_\mathrm{a}(\theta)-p_\mathrm{n}(\theta)|/p_\mathrm{n}(\theta)$, in logarithmic scale. Names of the lines are the same as in Table \ref{tab:p(theta)}.}
\label{fig:error}
\end{figure}

In Figure \ref{fig:p(theta)} we see that the analytical expression given by Eq. (\ref{fig:p(theta)}) rapidly converges to the numerical solution at $\theta\gg 1$, much slower converges to it at $\theta\ll 1$, and deviates from the numerical solution by $\sim 8\%$ at $\theta\sim 0.4$. Figure \ref{fig:error} shows the relative error of this approximation as a function of $\theta$. As we plot the absolute value of the relative error in logarithmic scale, the line tends to $-\infty$ whereever the numerical solution and the analytical approximation happen to coincide, which is responsible for a ``spiky'' appearance of the plot.

Led by the dream of obtaining a maximally precise analytical approximation, we try to improve upon Eq. (\ref{p_unified_analytic}) by keeping the first and the last terms in the denominator as they are to guarantee correct asymptotics, fine-tuning coefficients and power indices in the other terms, and maybe adding even more terms in the denominator. The results of several such tries are presented as analytical expressions in Table \ref{tab:p(theta)}, as $p(\theta)$ plots in Figure \ref{fig:p(theta)}, and as plots of the absolute value of the relative error in Figure \ref{fig:error}. We call different tries with the word ``fit'' followed by the number of free parameters and (if needed) a letter to distinguish between different fits with the same number of free parameters. The accuracy levels below our numerical error of 0.001 are shown in Figure \ref{fig:error} as a gray shaded area. It makes no sense to push the curves too deep into this area, as all curves in the gray region are dominated by the error in the input numerical data and thus virtually indistinguishable from the accuracy standpoint.

To keep track of the competition between the different fits, we use the root mean square relative error of each method, averaged over a logarithmic range of $\theta$, $-3<\lg(\theta)<3$:
\begin{equation}
\mathrm{RMSE} = \frac{1}{6}\int\limits_{-3}^{3}\frac{(p_\mathrm{a}(\theta)-p_\mathrm{n}(\theta))^2}{p_\mathrm{n}(\theta)^2}d\lg{\theta}.
\end{equation}
Here, $p_\mathrm{a}$ and $p_\mathrm{n}$ are the analytical and numerical expressions for the Yarkovsky pressure. The integration is performed over $\lg{\theta}$ and not over $\theta$ to avoid giving too much weight to large values of $\theta$. The prefactor $6=\lg(1000)-\lg(0.001)$ in the denominator is the width (in the logarithmic scale) of the range of $\theta$, over which the averaging is performed. 

We use Figure \ref{fig:competition} as the main jousting arena between different candidate approximations. An ideal approximation (a) should have the least possible RMSE and (b) ideally would be as simple as possible and thus have less free parameters. From the plot we can see, that, as usual, accuracy can be traded for simplicity: the more free parameters an expression has, the more precise it is. Depending on the preferred simplicity and accuracy, different expressions from Table \ref{tab:p(theta)} can be chosen. In general, it is not worth it to aim at the RMSE far below 0.001 (gray shaded area), as the numerical solution itself is not precise, and 0.001 was our targeted numerical error.

Our approximation of choice is the one with 6 free parameters, and thus 5 summands in the denominator. It is asymptotically correct at $\theta\rightarrow 0$ and $\theta\rightarrow \infty$, and it has RMSE = 0.000639 in the range $-3<\lg(\theta)<3$. Its maximal relative error in this range is attained at $\theta\approx 3$ and equals $|p_\mathrm{a}-p_\mathrm{n}|/p_\mathrm{n}\approx 0.0013$. This approximation incapsulates the main result of this article, and will be used as the basis for other parametric studies in the next articles of the series.

\section{Latitude dependence of the Yarkovsky pressure}
\label{sec:latitude_dependence}
Now, consider a non-equatorial surface element with an arbitrary asteroid latitude $\psi$. It creates the Yarkovsky force per unit area equal to
\begin{equation}
\frac{dF}{dS}=P_0 p_{\theta,\psi}(\theta,\psi),
\label{p_theta_psi-def}
\end{equation}
where $P_0$ is defined by Eq. (\ref{P0-def}), whereas $p_{\theta,\psi}(\theta,\psi)$ is some non-dimensional function of the thermal parameter and latitude. At the equator ($\psi=0$) this function is the same as the previously defined $p_\theta$: $p_{\theta,\psi}(\theta,0)=p_\theta(\theta)$. In the toy model of Eq. (\ref{FYark-initial}), we assumed that $p_{\theta,\psi}(\theta,\psi)=p_\theta(\theta)\cos^2\psi$, and this resulted in $p_s^d=1/\pi$ for a spherical asteroid. As we will see now, the true expression for $p_{\theta,\psi}(\theta,\psi)$ is more complicated, resulting in additional factors in Eq. (\ref{general-simple}) even for a spherical asteroid.

To derive these factors, we will revise the equations from the previous section and add some coefficients $\cos\psi$ where they are needed. The previous section was equally true for the diurnal Yarkovsky produced by an equatorial point of an asteroid with $\varepsilon=0^\circ$ or $180^\circ$ and for the seasonal Yarkovsky produced by a polar point of an asteroid with $\varepsilon=90^\circ$. Still, now we will restrict ourselves to the diurnal Yarkovsky effect. So, let the amendment of the equations begin.

First, $F_\mathrm{proj}$ in Eq. (\ref{P_dimensional}) gets multiplied by $\cos\psi$, which takes care of the projection of the Yarkovsky force at arbitrary latitude onto the orbit and turns the expression for the non-dimensional pressure into
\begin{equation}
p_{\theta,\psi} = \frac{2}{3} \langle \tau^4\big|_{z=0}\sin{\phi}\rangle_\phi\cos\psi.
\label{p_non-dimensional_psi}
\end{equation}
Next, the set of equations Eqs. (\ref{pde})-(\ref{boundary_condition3}) remains the same, and only the non-dimensional $\alpha(\phi)$ is multiplied by the factor $\cos\psi$.

The easiest way to proceed further is to note, that Eqs. (\ref{pde})-(\ref{boundary_condition3}) turn into themselves after the transformation
$\theta\rightarrow\theta\cos{^{3/4}\psi}$,
$\tau\rightarrow\tau\cos{^{1/4}\psi}$, $\alpha\rightarrow\alpha\cos\psi$.
Then Eq. \ref{p_non-dimensional_psi} for $p(\theta,\psi)$ assumes the form
\begin{equation}
p_{\theta,\psi}(\theta\cos{^{3/4}\psi},\psi)=\frac{2}{3} \langle (\tau\cos{^{1/4}\psi})^4\big|_{z=0}\sin{\phi}\rangle_\phi\cos\psi.
\end{equation}
Comparing it with Eq. (\ref{p_non-dimensional}) for previously defined pressure at the equator $p(\theta,0)$, we get
\begin{equation}
p_{\theta,\psi}(\theta\cos{^{3/4}\psi},\psi)=p_\theta(\theta)\cos{^2\psi}.
\end{equation}
Finally, substituting $\theta\cos{^{-3/4}\psi}$ for $\theta$, we get the general expression for $p$ at any latitude in terms of $p$ at the equator,
\begin{equation}
p_{\theta,\psi}(\theta,\psi)=p_\theta(\theta\cos{^{-3/4}\psi})\cos{^2\psi}.
\label{psi_transformation}
\end{equation}

Thus all our expressions obtained for the diurnal Yarkovsky effect at the equator can be generalized for arbitrary latitude. For example, Eq. (\ref{p_unified_analytic}) transforms into
\begin{equation}
p_{\theta,\psi}=\frac{\theta}{{\displaystyle\frac{5.006\,\theta^2}{(\cos\psi)^{11/4}}}+{\displaystyle\frac{12\,\theta}{(\cos\psi)^{2}}}+{\displaystyle\frac{6.367\,\theta^{1/4}}{(\cos\psi)^{23/16}}}+{\displaystyle\frac{7.159}{(\cos\psi)^{5/4}}}}.
\label{p_unified_analytic_psi}
\end{equation}
(Here, we have substituted the numerical values of the coefficients in order to both have this form of equation written explicitly somewhere in the article and because otherwise we couldn't fit the fraction into single line.)

\begin{figure}
\begin{center}
\includegraphics[width=.49\textwidth]{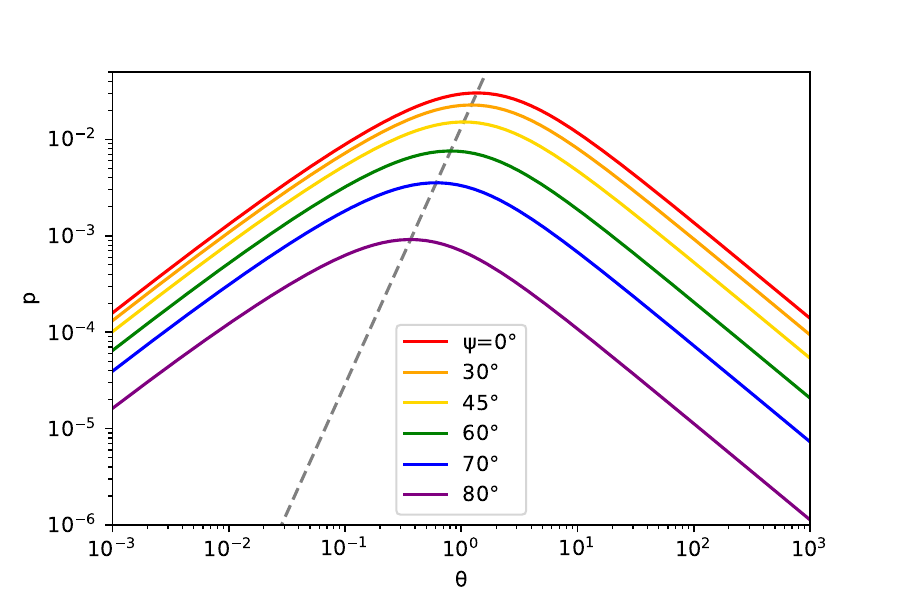}
\caption{The non-dimensional Yarkovsky pressure $p$ as a function of the thermal parameter $\theta$ at different latitudes $\psi$.}
\label{fig:p(theta,psi)}
\end{center}
\end{figure}

In Figure \ref{fig:p(theta,psi)} we plot $p_{\theta,\psi}$ as a function of $\theta$ at different latitudes $\psi$ with lines of different colors. We see that at all the latitudes the lines have similar shapes, just as Eq. (\ref{psi_transformation}) prescribes. At higher latitudes, the lines shift down and to the left because of the factors $\cos{^{-3/4}\psi}$ and $\cos{^2\psi}$ respectively in Eq. (\ref{psi_transformation}).

\section{Integration of the Yarkovsky pressure over a spherical asteroid}
\label{sec:spherical_ast_integr}
Now let us bring the theory one step closer to the practical applications for real objects, and integrate the Yarkovsky pressure over the surface of a spherical asteroid. Such a model is indeed the model of choice for objects whose shapes are unknown. The Yarkovsky effect for asteroids of more complex shapes will be the focus of the next paper of the series.

Using the notations from Eq. (\ref{p_theta_psi-def}), the Yarkovsky force integrated over the surface of a spherical asteroid with the radius $R$ and zero obliquity is
\begin{equation}
F=\int dF =P_0 R^2 \cdot 2\pi\int\limits_{-\pi/2}^{\pi/2}p_{\theta,\psi}(\theta,\psi)\cos{\psi}d\psi
\label{F_def}
\end{equation}

We already used the same reasoning when deriving Eq. (\ref{FYark-initial}) from a toy model in the beginning of the article, but that time we had no expression for $p_{\theta,\psi}(\theta,\psi)$ and used simplistic physical arguments to assume that $p_{\theta,\psi}(\theta,\psi)\propto\cos^2\psi$. This resulted in the expression is similar to Eq. (A.3) in \cite{kyrylenko2021}, with the coefficient 0.033 in the latter being roughly equivalent to the maximum in the numerical curve in Figure \ref{fig:p(theta)}. Still, from Eq. (\ref{psi_transformation}) we see that $p_{\theta,\psi}(\theta,\psi)\propto\cos^2\psi$ only if $p_{\theta,\psi}(\theta,\psi)$ is independent on its first argument, i.e. close to its maximum. Thus this approach can be more or less true at $\theta\sim 1$, but should break at $\theta\ll 1$ or $\theta\gg 1$, where the dependence of $p$ on $\theta$ is more prominent.

By comparing Eq. (\ref{p_unified_analytic_psi}) to Eq. (\ref{p_unified_analytic}) one can see that
$p_{\theta,\psi}(\theta,\psi)\approx p_\theta(\theta)\,(\cos\psi)^{11/4}$ for $\theta\gg 1$ and $p_{\theta,\psi}(\theta,\psi)\approx p_\theta(\theta)\,(\cos\psi)^{5/4}$ for $\theta\ll 1$.
In these two limiting cases, the integral in Eq. (\ref{F_def}) equals
\begin{align}
F=F_0\frac{3\sqrt{\pi}\Gamma\left(\frac{19}{8}\right)}{4\Gamma\left(\frac{23}{8}\right)}=0.90887 F_0,\,\,\,\theta\gg 1,\nonumber\\
F=F_0\frac{3\sqrt{\pi}\Gamma\left(\frac{13}{8}\right)}{4\Gamma\left(\frac{17}{8}\right)}=1.1250 F_0,\,\,\,\theta\ll 1,
\end{align}
where $F_0=\frac{8\pi}{3}P_0 R^2 p_\theta(\theta)$ is the force from the simplified model of Eq. (\ref{FYark-initial}).

We can introduce the new non-dimensional function defined as $p_\mathrm{\theta,sph}(\theta)=p_\theta(\theta)F/F_0$. It does not have such a clear physical meaning as $p_\theta(\theta)$ had. It is no longer a non-dimensionalized Yarkovsky force at the equator. It also includes a factor $F/F_0$ that corrects for the difference between the exact surface integration and the toy model. As a part of surface integration, this factor could have been incorporated into $p_s(\mathrm{shape})$. But as it depends on $\theta$, we prefer to incorporate it into $p_\theta(\theta)$, even at the cost of making the physical meaning of $p_\mathrm{\theta,sph}(\theta)$ less clear than it was for $p_\theta(\theta)$. This newly introduced $p_\mathrm{\theta,sph}(\theta)$ is convenient as its substitution into Eq. (\ref{general-simple}) corrects for the different dependence on $\psi$ at different $\theta$ (at least for a spherical asteroid) and produces a more precise result than the toy model did.

\begin{figure}
\begin{center}
\includegraphics[width=.49\textwidth]{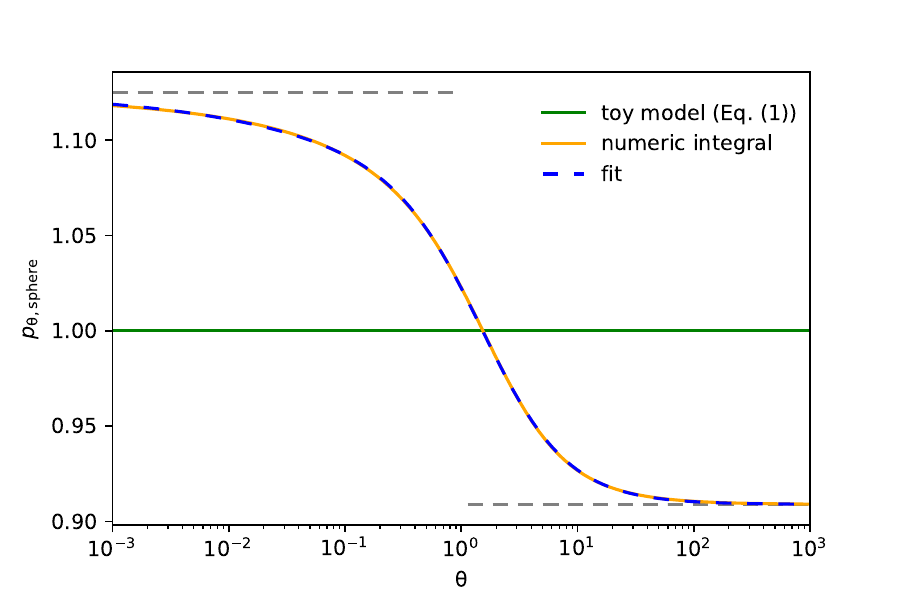}
\caption{The shape factor $p_\mathrm{\theta,sph}(\theta)/p_\theta(\theta)$ for a spherical asteroid as a function of the thermal parameter $\theta$.}
\label{fig:sphere_fraction}
\end{center}
\end{figure}

In Figure \ref{fig:sphere_fraction} we plot the ratio $p_\mathrm{\theta,sph}(\theta)/p_\theta(\theta)$ as a function of $\theta$. The range between the two limiting values 0.9089 and 1.1250 is shown as a blue shaded area, whereas the green horizontal line show the value $1$ that corresponds to the intermediate power index of 2. The orange curve is the result of numerical integration of Eq. (\ref{F_def}). We see that it starts at the upper boundary of the blue area at $\theta\ll 1$ and smoothly transfers to its lower boundary at $\theta\gg 1$. Even if one uses the approximate value 2 of the power index throughout the entire range of $\theta$, it introduces the error below 15\%, which is still much smaller than the 100\% introduced by the ``standard'' approximation. 

A good fit to the plot is given by the following analytical expression:
\begin{equation}
\frac{p_\mathrm{\theta,sph}(\theta)}{p_\theta(\theta)}=\frac{0.9089\cdot\theta^{-0.2737} + 0.3333 + 1.1250\cdot\theta^{0.8726}}{\theta^{-0.2737} + 0.3309 + \theta^{0.8726}}
\label{knee}
\end{equation}
In Figure \ref{fig:sphere_fraction}, this fit is marked with the blue dashed line. One can see that the fit is so good that it is virtually indistinguishable from the exact numerical result shown with the orange line.

\begin{figure}
\begin{center}
\includegraphics[width=.49\textwidth]{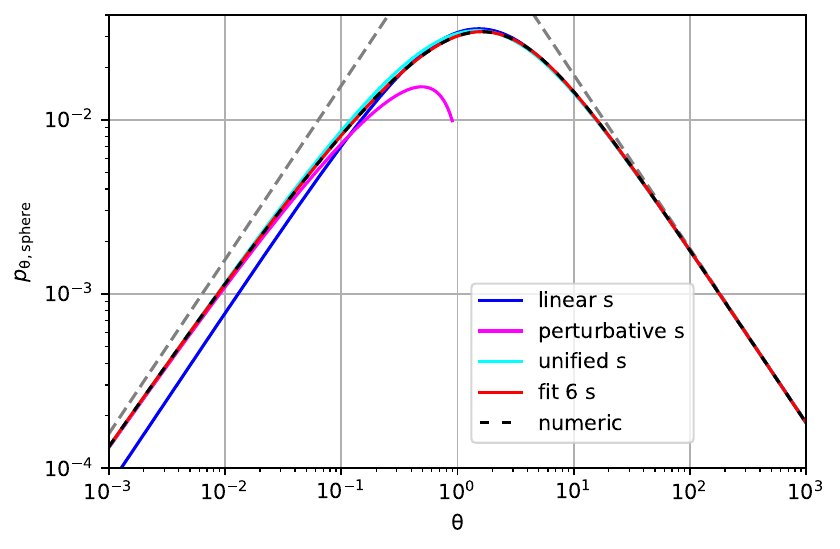}
\caption{$p_\mathrm{\theta,sph}(\theta)$ for a spherical asteroid as a function of the thermal parameter $\theta$.}
\label{fig:sphere}
\end{center}
\end{figure}

\begin{figure}
\begin{center}
\includegraphics[width=.49\textwidth]{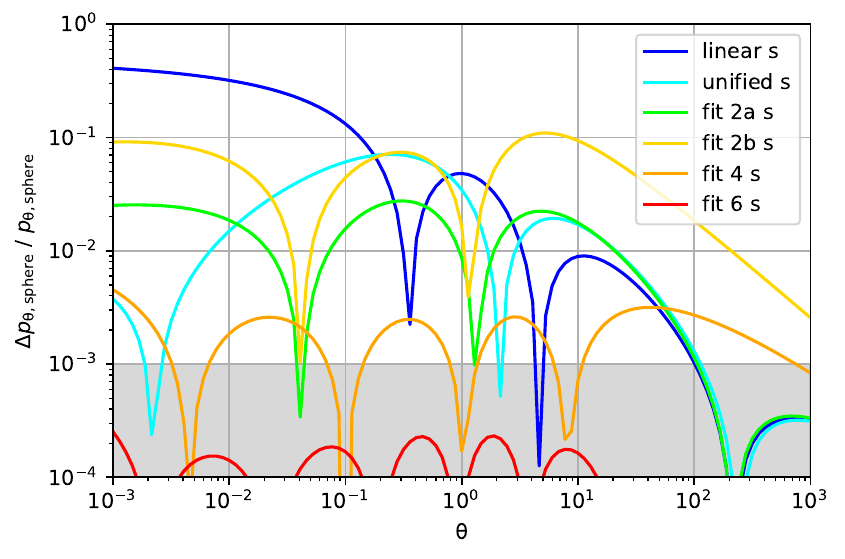}
\caption{Same as Fig. \ref{fig:error}, but with the error of the Yarkovsky effect integrated over a sphere instead of the Yarkovsky effect at an equatorial point. Names of the lines are the same as in Table \ref{tab:p_sph(theta)}.}
\label{fig:sphere_error}
\end{center}
\end{figure}

Still, it is impractical to use a fitted analytical expression for $p_\theta(\theta)$, and then to multiply it by another fitted analytical expression for $p_\mathrm{\theta,sph}(\theta)/p_\theta(\theta)$, thus getting a cumbersome final expression for $p_\mathrm{\theta,sph}(\theta)$. Instead, it is more straightforward to fit the numerical data for $p_\mathrm{\theta,sph}(\theta)$ as they are. Given that $p_\mathrm{\theta,sph}(\theta)/p_\theta(\theta)\approx 1$, it is natural to use for $p_\mathrm{\theta,sph}(\theta)$ the expressions, similar in form to the ones we used previously for $p_\theta(\theta)$. We present the best-fit expressions in Table \ref{tab:p_sph(theta)}, show the corresponding plots in Figure \ref{fig:sphere} (which is similar to Figure \ref{fig:p(theta)}), and the relative error of the fit in Figure \ref{fig:sphere_error} (which is for the best-fitting expression even smaller than the error in Figure \ref{fig:error}).

\section{Results}
\label{sec:results}
Accumulating the wisdom gained throughout the article in one place, we obtain the final expression for the Yarkovsky effect, acting on a spherical asteroid of the radius $R$ with the albedo $A$ and obliquity $\varepsilon$ at the orbit with the semimajor axis $a$ and eccentricity $e$:
\begin{equation}
\Dot{a}=\frac{(1-A)L_{\mathrm{\odot}}}{ Rc\rho\sqrt{M_{\mathrm{\odot}}Ga}}\left(\frac{p_{\theta,\mathrm{sph}}(\theta)\cos\varepsilon}{\pi(1-e^2)}-\frac{p_{\theta,\mathrm{sph}}(\theta_s)\sin^2\varepsilon}{2\pi(1-e^2)}\right),
\label{res4}
\end{equation}
the first term in the brackets corresponding to the diurnal Yarkovsky effect, and the second to the seasonal effect.
For the non-dimensional Yarkovsky pressure $p_{\theta,\mathrm{sph}}(\theta)$ we recommend the following equation regardless of whether diurnal or seasonal effect is considered:
\begin{equation}
p_{\theta,\mathrm{sph}}(\theta)=\displaystyle\frac{\theta}{a_p+x\theta^{\xi}+y\theta^{\eta}+z\theta^{\zeta}+d_l\theta^2}
\label{res2}
\end{equation}
with the coefficients $x=6.9138$, $\xi=0.25260$, $y=8.7966$, $\eta=0.66496$, $z=5.8895$, $\zeta=1.1225$, $a_p=-\frac{45\sqrt{\pi}\Gamma(-3/8)}{4\sqrt{2}\Gamma(1/8)}\approx 7.1592$, $d_l=\frac{3}{\sqrt{2}}\pi^{3/4}\approx 5.0057$. Still, other equations can be found in Table \ref{tab:p_sph(theta)}. Alternatively, readers can conduct their own more detailed numerical simulations, repeating the authors' steps described in this article, and thus refine the coefficients in the approximation function.

The diurnal and seasonal thermal parameters, entering Eq. (\ref{res4}) are expressed as
\begin{equation}
\begin{aligned}
\theta&=&\frac{(C\rho\kappa\omega)^{1/2}}{(\epsilon\sigma)^{1/4}}\left(\frac{4\pi a^2}{(1-A)L_\odot}\right)^{3/4},\\
\theta_s&=&\frac{(C\rho\kappa\omega_s)^{1/2}}{(\epsilon\sigma)^{1/4}}\left(\frac{4\pi a^2}{(1-A)L_\odot}\right)^{3/4},
\label{res3}
\end{aligned}
\end{equation}
where $\omega$ and $\omega_s$ are the mean angular velocities of diurnal and seasonal rotation, whereas $C$, $\rho$, $\kappa$, and $\epsilon$ are the heat capacity, density, thermal conductivity, and emissivity of the asteroid soil.

Equation (\ref{res4}) is more precise than all previous expressions for the Yarkovsky effect due to correct analytical consideration of the limiting case of small thermal parameters and fitting of high-accuracy numerical results at intermediate thermal parameters. The general perturbative approach for the case of small thermal parameters is an important result of the article \textit{per se}, although its further justification is required.

The obtained formula for the Yarkovsky effect only approximately accounts for the orbit eccentricity and the asteroid obliquity, which will be corrected in the following articles, and the accuracy of Eq. (\ref{res4}) in this respect is still to be investigated.

What we have solidly established here, on the other hand, is the following expression for the Yarkovsky drift of a spherical asteroid with zero obliquity on a circular orbit:
\begin{equation}
\Dot{a}=\frac{(1-A)L_{\mathrm{\odot}}p_{\theta,sph}(\theta)}{\pi Rc\rho\sqrt{M_{\mathrm{\odot}}Ga}}.
\label{res1}
\end{equation}
This expression is correct at the accuracy of $10^{-3}$ in the range $10^{-3}<\theta<10^{3}$, as it can be seen in Figure \ref{fig:sphere_error}. By connecting $\Dot{a}$ with the Yarkovsky force, or by directly applying Eq. (\ref{FYark-initial}), we get the following expression for the Yarkovsky force:
\begin{equation}
F=\frac{2(1-A)L_{\mathrm{\odot}} R^2}{3cr^2}.
\end{equation}
This force acts perpendicularly to the radius-vector drawn from the Sun to the asteroid, whose length we denote $r$. We recommend this expression for use in numerical models of asteroid dynamics, as well as for further theoretical analysis, e.g. for rigorous integration of this force's torque over an elliptic asteroid orbit.

\section{Discussion}
\label{sec:discussion}

The application of our new, more precise semi-analytical formula (\ref{res4}) for the Yarkovsky drift, instead of traditional formulas based on linear theory, may affect various other branches of asteroid science that build upon it. However, we first demonstrate that the majority of asteroids indeed lie outside the range of applicability of the ``standard'' linear theory.

%To make connection with the observables, let us express the thermal parameter $\theta$ in terms of the surface thermal inertia $\Gamma=\sqrt{\kappa C\rho}$, the rotational period $P_\mathrm{rot}=2\pi/\omega$ and the mean distance from the Sun (semimajor axis) $a$:
%\begin{equation}
%\theta=\frac{\Gamma}{(\epsilon\sigma)^{0.25}}\left(\frac{2\pi}{P_\mathrm{rot}}\right)^{0.5}\left(\frac{4\pi a^2}{(1-A)L_\odot}\right)^{0.75}
%\label{theta_phys_meaning}
%\end{equation}

\begin{figure*}
\centering
\includegraphics[width=0.493\textwidth]{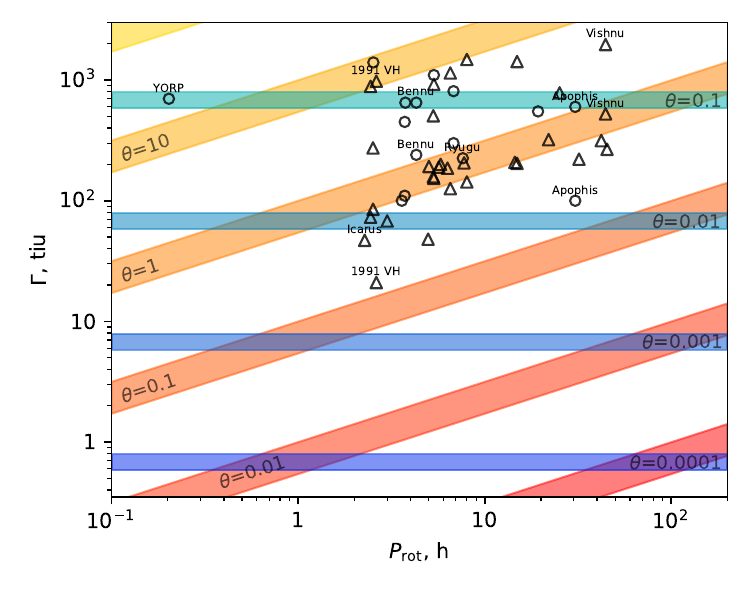}
\includegraphics[width=0.49\textwidth]{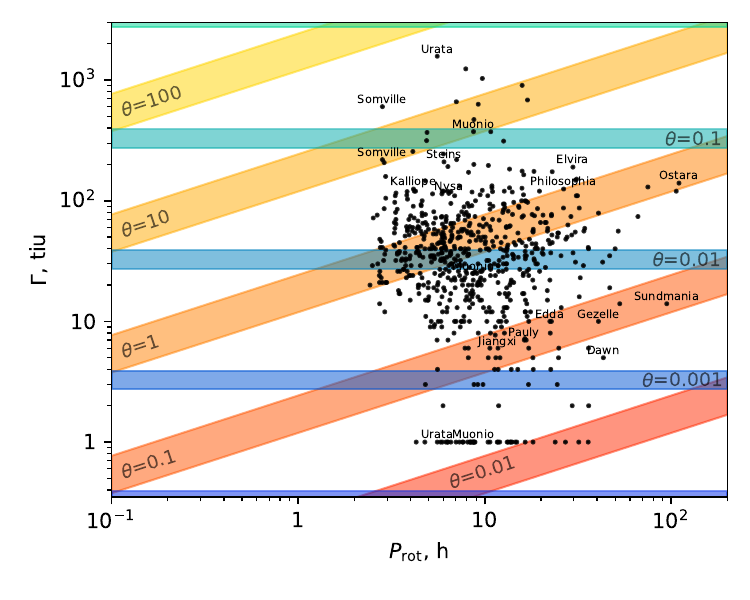}
\caption{Dependence of the thermal parameters $\theta$ on the rotation periods $P$ (in hours) and thermal inertias $\Gamma$ (in TIU, thermal inertia units,
TIU = J m$^{-2}$ K$^{-1}$ s$^{-1/2}$) for the near-Earth asteroids (left, assumed distance to the Sun $a=1-1.5$ au) and the main-belt asteroids (right, assumed distance to the Sun $a=2.05-3.3$ au). The other parameters are assumed to be $A=0.15$ and $\epsilon=0.9$. The plotted asteroids are taken from \cite{maclennan2021}, \cite{hung_etal2022}, and \cite{choukroun_etal2025}.}
\label{fig:thetas-obs}
\end{figure*}

As we can see in Figures \ref{fig:p(theta)}--\ref{fig:error} and \ref{fig:sphere}--\ref{fig:sphere_error} the linear theory overestimates the Yarkovsky effect by $\sim 10\%$ at intermediate thermal parameters $\theta\approx 1$ and dramatically underestimates it at $\theta<0.1$, ultimately reaching an approximately twofold error at small $\theta$. In Figure \ref{fig:thetas-obs} we can see, for which asteroid which approximation works. The rotation periods $P_\mathrm{rot}=2\pi/\omega$ of asteroids are plotted along the $x$-axis, and the thermal inertias $\Gamma=\sqrt{\kappa C\rho}$ along the $y$-axis. Individual asteroids of differing diameters with measured thermal inertias are shown with points \citep{maclennan2021, hung_etal2022, choukroun_etal2025}. The colorful stripes on the plot show the corresponding thermal parameters, computed using Eq. (\ref{res3}).
The left-hand panel shows $P_\mathrm{rot}$ and $\Gamma$ for the near-Earth asteroids, and the corresponding thermal parameters $\theta$ are computed assuming $\epsilon=0.9$, $A=0.15$, and $a$ ranging from 1 to 1.5 au. The right-hand panel shows the main-belt asteroids assuming $a$ ranging from 2.05 to 3.3 au. As $a$ covers a broad range, the fixed values of $\theta$ correspond not to thin lines but to broad stripes in the plot, with the upper bound of $a$ corresponding to the lower boundary of each stripe, and the lower bound of $a$ -- to the upper boundary of the stripe. These stripes indicate the approximate values of the diurnal (red--yellow colors) and seasonal (blue--green colors) thermal parameters.

\begin{figure*}
\centering
\includegraphics[width=0.493\textwidth]{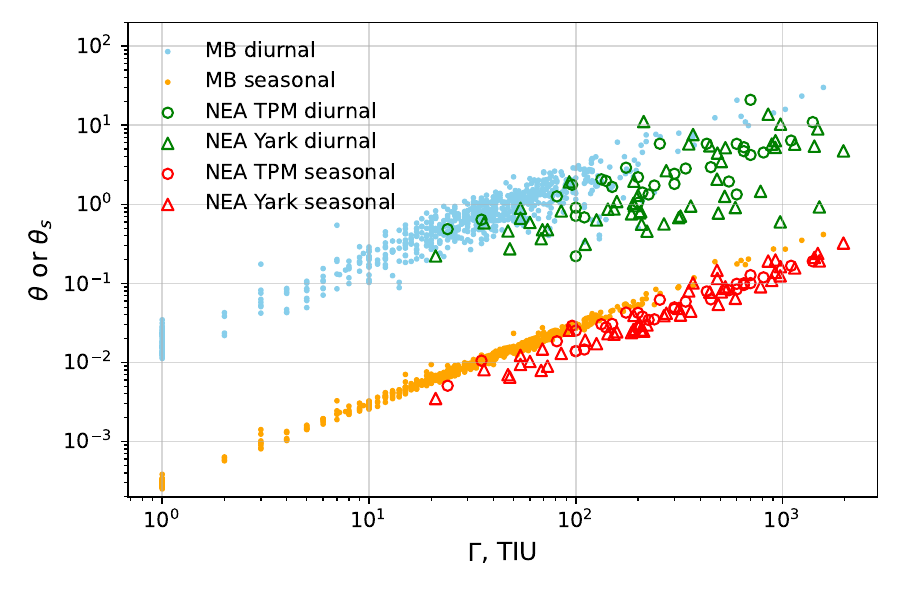}
\includegraphics[width=0.49\textwidth]{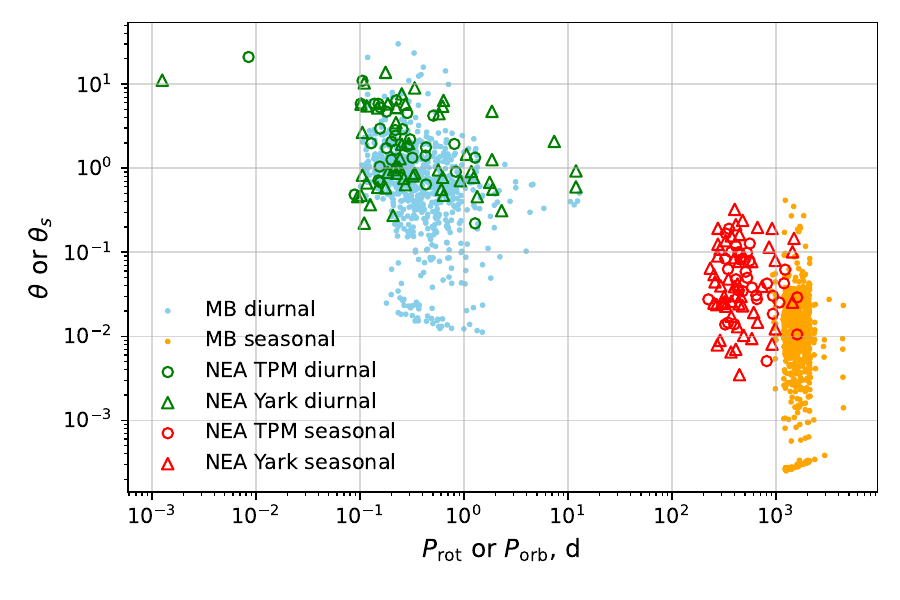}
\caption{Dependence of the thermal parameters $\theta$ and $\theta_s$ on the thermal inertia $\Gamma$ (left) and the rotation period $P_\mathrm{rot}$ or the orbital period $P_\mathrm{orb}$ (right). The plotted asteroids are taken from \cite{maclennan2021}, \cite{hung_etal2022}, and \cite{choukroun_etal2025}.}
\label{fig:thetas-obs2}
\end{figure*}

Another representation of the same data on the thermal parameters of asteroids is shown in Figure \ref{fig:thetas-obs2}. Here, the diurnal $\theta$ and seasonal $\theta_s$ thermal parameters of each asteroid are plotted along the vertical axis, whereas the horizontal axis shows the thermal inertia $\Gamma$ (left-hand panel) or the rotation period (of axial and orbital rotation for the diurnal and seasonal Yarkovsky effect correspondingly; right-hand panel).

The two analytical theories that we discuss in the article are valid for $\theta \ll 1$ and $\theta \gg 1$, whereas in Figures \ref{fig:thetas-obs} and \ref{fig:thetas-obs2} we see that most asteroids have $\theta\sim 1$. This can be partially attributed to observational bias, as $\theta\sim 1$ are the easiest for the observational measurement. (54509) YORP is the only asteroid that lies in the area $\theta \gg 1$, and for which the linear theory is well applicable.

Large values of the thermal parameter are possible only with sufficiently fast rotation of the object and high thermal inertia of its surface, especially if this object is located at great distance from the Sun \citep{spencer1989}. Thus, we can safely state that the ``standard'' linear theory is still applicable for most meteoroids and several-meter-sized asteroids. In other cases, we recommend applying our new semi-analytical formula (\ref{res4}).

Improvements in the analytical modeling of the Yarkovsky effect directly affect a wide range of important problems in the physics and dynamics of small Solar System bodies.
In particular, determining an asteroid's thermal inertia $\Gamma$ from its measured Yarkovsky drift poses a compelling inverse problem. This methodology, along with its rigorous testing, was implemented for various near-Earth asteroids by \cite{fenucci_etal2021, fenucci_etal2023} and \cite{novakovic_etal2024}.
Figure \ref{fig:sphere} demonstrates that solving this problem typically yields two distinct solutions for the thermal parameter and, consequently, two values of the thermal inertia. Still, it is not always clear which of the two $\Gamma$ values is physically relevant. \cite{novakovic_etal2024} showed that for many NEAs, both solutions of the thermal inertia are entirely plausible.
Crucially, our revised theory of the Yarkovsky effect suggests that the lower bound of $\Gamma$ may be overestimated if the thermal parameter $\theta$ falls below 0.1. 
The aforementioned studies also call into question the validity of low thermal inertia estimates for rapidly rotating asteroids. 
While the mechanisms and processes leading to such low $\Gamma$ values are beyond the scope of the present study, such values are consistent and expected near the maximum Yarkovsky drift ($\theta\sim 1$) for both NEAs and main-belt asteroids, as seen from Figures \ref{fig:thetas-obs} and \ref{fig:thetas-obs2}.
 
As for the other inverse problems, our correction of the Yarkovsky expression should correspondingly change the Yarkovsky-based asteroid density estimates in cases where the asteroid shapes are poorly restricted, no full-fledged thermal modeling is conducted, and the density is evaluated from the approximate analytical formulas. Similarly, when the Yarkovsky effect is used for asteroid family dating, our new expression will affect the family age estimate (the age will increase by several per cent for families with $\theta\sim 1$), although this change might be smaller than other sources of uncertainty (such as poorly restricted asteroid densities, shapes, thermal inertias, rotation rates, and obliquities).

The suggested increase of accuracy of the Yarkovsky drift evaluation must be particularly important for predicting collisions of potentially hazardous asteroids (PHAs) with Earth. Such predictions are usually limited by a time span of about 100 years\footnote{\url{https://cneos.jpl.nasa.gov/sentry/}}, and the Yarkovsky effect is often the major source of uncertainty that prohibits precise orbital simulations into a distant future. Although for strongly chaotic orbits the increase of the Yarkovsky effect accuracy can increase the reliably simulated time span only logarithmically, even a logarithmic increase is highly demanded for such a vital problem as planetary defense. Still, for now our theory is precise only for a spherical asteroid with zero obliquity at a circular orbit, while the non-zero obliquity, eccentricity, and seasonal component are included only in the simplest heuristic manner, and a more rigorous theoretical model is needed to prevent the Yarkovsky effect from dominating the error in asteroid hazard prediction.

In Appendix \ref{sec:app:assumptions}, we do our best to outline the theory's most important limitations, in order to warn the readers against using our model outside its range of applicability and to urge them to generalize the proposed theory by getting rid of these limitations. For our part, we plan to devote the next articles of this series to integration of the Yarkovsky force over the asteroid of a complex shape, investigation of the effect of obliquity and eccentricity, and evaluation of the seasonal Yarkovsky effect at the equally high accuracy as the results for a spherical asteroid at zero obliquity and zero eccentricity presented in this article.

\section{Conclusion}

The main results of this article are several expressions for the Yarkovsky effect. We derive a new theory for the Yarkovsky effect valid for $\theta\ll 1$, smoothly sew it with the standard linear theory valid for $\theta\gg 1$, and fit the parameters in the resulting expression to get the best consistency with the results of the numerical simulation. The resulting semi-analytical expression for the Yarkovsky effect is correct for all values of $\theta$ with the relative error of about $10^{-3}$. It contrasts with the ``standard'' widely used linear model, whose application results in an error factor of 2 for small thermal parameters. We present the model for a separate surface element and for a spherical asteroid. It is most interesting to integrate the Yarkovsky force over the complex shape of an asteroid and thus to preserve the accuracy of the presented theory from being compromised by the errors of a loose treatment of the shape. But let this question be a cliffhanger for our next article.

\begin{acknowledgements}
%This work was partially funded by the National Research Foundation of Ukraine, project N2020.02/0371 ``Metallic asteroids: search for parent bodies of iron meteorites, sources of extraterrestrial resources''. 
The authors express their gratitude to all those people who defend Ukraine from Russian aggression and thanks to whom it became possible for us to prepare this article.
OG is grateful to the European Federation of Academies of Sciences and Humanities (ALLEA EFDS-FL-16). VL acknowledges financial support from the German Excellence Strategy via the Heidelberg Cluster of Excellence (EXC 2181 - 390900948) ``STRUCTURES''. She also thanks for computing resources provided by the Ministry of Science, Research and the Arts (MWK) of the State of Baden-W\"{u}rttemberg through bwHPC and the German Science Foundation (DFG) through grants INST 35/1134-1 FUGG and 35/1597-1 FUGG, and also for data storage at SDS@hd funded through grants INST 35/1314-1 FUGG and INST 35/1503-1 FUGG.

\end{acknowledgements}

\appendix

\section{Assumptions and limitations}
\label{sec:app:assumptions}
To clarify the boundaries of the proposed theory and set the floor for the following articles of this series, let us clearly verbalize the most important assumptions and limitations of our approach.

First of all, there was a set of assumptions outlined in Section \ref{sec:scope}. We neglected the higher-order terms in the Fourier series on obliquity in Eqs. (\ref{eps-series-odd}) and (\ref{eps-series-even}). In Eq. (\ref{F-dot-v}), we neglected the dependence of the thermal parameter on the distance to the Sun. In this equation, we also averaged the Yarkovsky force over the asteroid's orbit, which is a valid description of the asteroid's long-term evolution, but may be insufficient to accurately describe its orbit on a month-to-month basis. The possible dependence of the Yarkovsky effect on the
angle $\gamma$ between the perihelion and the asteroid equinox point in Eq. (\ref{yarkovsky-almost-the-most-general}), the imperfect factorization of the force into the product of terms depending on separate variables, and the possible mixing between the diurnal and seasonal Yarkovsky effects are but a few assumptions that we have outlined before and whose more considerate investigation we postpone for later. But let us now more carefully list the assumptions that we have implicitly assumed in this paper and that we want to keep throughout most of this series of articles. Going beyond each of these assumptions is worth another series of papers, and in some cases such papers have already been written by other authors. 

Equation (\ref{P-inst}) assumes the Lambert heat emission law. It is known to be a good approximation for some surfaces, but the surfaces with substantially non-Lambertian indicatrix can produce a coefficient different from 2/3 in Eq. (\ref{P-inst}). For example, the macroscopic features and rough surface of a real asteroid will cause effects such as the thermal-infrared beaming, which can increase the Yarkovsky orbital drift by tens of percent, and in some cases, even double it. (For more details, see \cite{rozitis_green2012}.)

The heat conduction equation Eq. (\ref{conductiv}) assumes that the heat capacity, heat conduction, and density are independent of the temperature and depth. In reality, the properties of the asteroid soil can strongly depend on the depth, and the heat conductivity is known to be a function of temperature (the most importantly, due to the radiative heat transfer). Moreover, the mere continuum approach embedded in Eq. (\ref{conductiv}) is just an approximation, which ignores the complex granular structure of the asteroid soil. If the number of grains in a volume of interest is limited, it puts the meaning of the average temperature $T$ under question, as the temperature can be different between different grains as well as within each grain. Further, the non-local radiation heat transfer between grains can force us to reject the differential heat conduction law Eq. (\ref{conductiv}) in favor of a much more complicated set of heat conduction equations for individual particles with the boundary conditions dictated by their mechanical contact and mutual illumination.

In addition, the surface layer bulk density $\rho$ from Eq. (\ref{conductiv}) can be equal to the bulk density $\rho$ of the entire asteroid (e.g. from Eq. (\ref{toy-final})) only in limited cases, whereas in general they can differ due to the possible presence of regolith on the surface of the asteroid, the processes of space weathering, mass differentiation etc.

The 1D heat conduction model expressed by Eq. (\ref{conductiv}) with the boundary condition at a large depth given by Eq. (\ref{boundary2}) ignores the curvature and roughness of the asteroid surface. Structures on the asteroid surface of the order of the thermal wavelength (such as boulders, trenches, etc.) could make the heat emission indicatrix asymmetric with respect to the normal of the surface (in a way similar to the tangential YORP effect). The heat conduction is also substantially non-1-dimensional in the case of small-sized asteroids, which, due to the through heat fluxes, rapidly acquire a near-uniform temperature distribution, which diminishes the Yarkovsky effect. 

The boundary condition expressed by Eq. (\ref{boundary}), in addition to the limitations mentioned in the previous paragraph, includes the thermal radiation law with the emissivity factor $\epsilon$, which is the average over the wavelengths, at which the radiation occurs. Therefore, $\epsilon$ can change, when temperature changes and Wien's displacement law pushes the spectral intensity distribution in and out of absorption bands of the asteroid surface material.

Bolometric Bond albedo $A$ in Eq. (\ref{E_of_alpha}) is assumed to be constant independently of the height of the Sun above the local horizon. This implies that the energy absorbed by the surface is directly proportional to the incoming solar energy flux projected onto the surface. This is not exactly true for realistic surfaces, where the fraction of the absorbed energy can depend on the incidence angle.

The simple expression for the geometric factor of the impinging solar power in Eq. (\ref{alpha}) or its generalization that takes into account the asteroid obliquity \citep{golubov2016} still ignores the effects of shadowing and self-illumination and thus is only approximately applicable to multiple systems (e.g. binaries) and non-convex asteroids, contact binaries, objects with craters or any other substantial relief or micro-relief of the surface. Furthermore, the rotation angle $\phi$ loses its simple physical meaning, and Eq. (\ref{alpha}) violates for tumbling asteroids. (For the numerical consideration of this problem in the special case of (99942) Apophis, see \cite{vokrouhlicky2015b_apophis}.)

One more limitation in Eq. (\ref{alpha}) is the assumption that the Sun is a point source. Accounting for the non-zero angular size of the Sun will result in a little bit slower change of illumination at the terminator.

The periodic initial condition Eq. (\ref{boundary3}) for the diurnal Yarkovsky effect can break if the axial rotation of the asteroid is so slow that during one axial period the asteroid shifts to a substantially different point of its orbit and thus experiences different illumination conditions (due to non-zero obliquity or eccentricity). The expression $\phi=\omega t$ with constant $\omega$ in Eq. (\ref{tau}) can also break for slow rotators, as $\omega$ has the sense of the synodic angular velocity, which can be non-constant due to the variation of the angular velocity of the asteroid around the Sun as a result of the ellipticity of its orbit. One more complication for slow rotators is the possibility of the aforementioned tumbling.  

Beyond this, we should always keep in mind that there is so much more to life than the Yarkovsky effect. In many cases, it can be outcompeted by the momentum change due to meteoroid bombardment, centrifugal mass shedding, sublimation of volatiles and other kinds of asteroid activity, as well as the Poynting--Robertson effect, solar radiation pressure etc.

\section{Different analytical approximations for the Yarkovsky effect}
\label{sec:app:approximations}
\begin{figure}
\centering
\includegraphics[width=\linewidth]{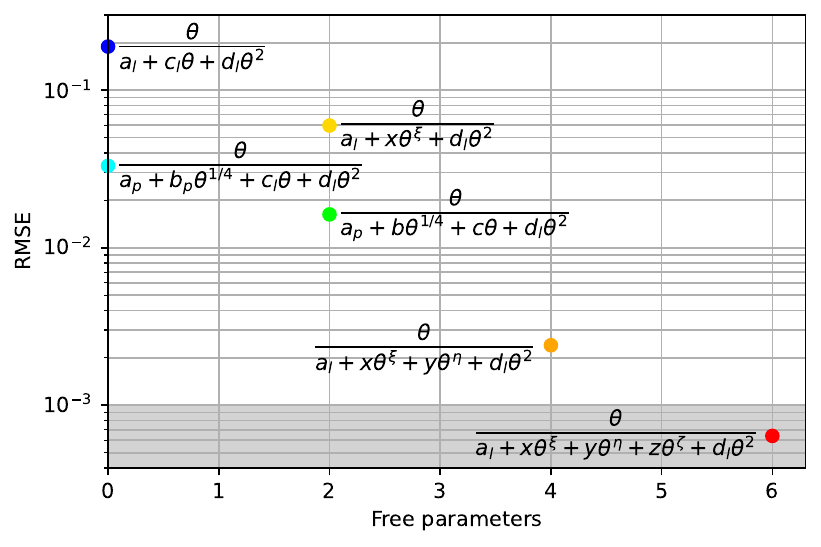}
\caption{Root mean square error in the range $-3<\lg(\theta)<3$ for different analytical fits plotted as a function of the number of free parameters of each fit. Values of the parameters in equations are given in Table \ref{tab:p(theta)}, colors  are the same as in Figure \ref{fig:error}.}
\label{fig:competition}
\end{figure}

In this Appendix, we assemble a more specialized information related to the analytical approaches used in our analysis. In Table \ref{tab:p(theta)} we present the best-fit expressions for the non-dimensional Yarkovsky pressure $p_\theta$ on an equatorial element, and in Figure \ref{fig:competition} we present their root mean square relative error in the range $-3<\lg(\theta)<3$ as a function of the number of free parameters.

Similarly, in Table \ref{tab:p_sph(theta)} we show the best-fit expressions for the non-dimensional Yarkovsky pressure $p_{\theta,\mathrm{sph}}$ for a spherical asteroid.

\begin{table*}
\caption{Different equations used for the non-dimensional Yarkovsky pressure $p_\theta$ throughout the article.}
\begin{center}
\renewcommand{\arraystretch}{3}
\begin{tabular}{p{2cm}|c|c|p{10cm}} 
\hline\textbf{}
Name & Equation & Range & Comments \\
\hline
linear & $\displaystyle\frac{\theta}{a_l+c_l\theta+d_l\theta^2}$ & $\theta\gg 1$ & analytical solution, Eq. (\ref{p_big_theta}) \\ 

perturbative asymptotics & $\displaystyle\frac{\theta}{a_p}$ & $\theta\ll 1$ & analytical solution, Eq. (\ref{p_small_theta}) \\

perturbative & $\displaystyle\frac{\theta}{a_p+b_p\theta^{1/4}}$ & $\theta\ll 1$ & analytical solution, Eq. (\ref{p_not_so_small_theta}) \\ 

unified & $\displaystyle\frac{\theta}{a_p+b_p\theta^{1/4}+c_l\theta+d_l\theta^2}$ & all $\theta$ & analytical equation constructed to fit both asymptotics, Eq. (\ref{p_unified_analytic}) \\ 

fit 2a & $\displaystyle\frac{\theta}{a_p+b\theta^{1/4}+c\theta+d_l\theta^2}$ & all $\theta$ & fit to numerical results; $b=8.4410$, $c=12.783$\\ 

fit 2b & $\displaystyle\frac{\theta}{a_p+x\theta^{\xi}+d_l\theta^2}$ & all $\theta$ & fit to numerical results; $x=22.464$, $\xi=0.52116$ \\ 

fit 4 & $\displaystyle\frac{\theta}{a_p+x\theta^{\xi}+y\theta^{\eta}+d_l\theta^2}$ & all $\theta$ & fit to numerical results; $x=10.076$, $\xi=0.30276$, $y=11.507$, $\eta=0.98217$ \\ 

fit 6 & $\displaystyle\frac{\theta}{a_p+x\theta^{\xi}+y\theta^{\eta}+z\theta^{\zeta}+d_l\theta^2}$ & all $\theta$ & fit to numerical results; $x=6.9138$, $\xi=0.25260$, $y=8.7966$, $\eta=0.66496$, $z=5.8895$, $\zeta=1.1225$\\ 
\hline
\end{tabular}
\end{center}
{\footnotesize Note: The equations use the following notations: $a_p=-\frac{45\sqrt{\pi}\Gamma(-3/8)}{4\sqrt{2}\Gamma(1/8)}\approx 7.1592$, $b_p=\frac{675\Gamma(-3/8)^2}{16\sqrt{2}\Gamma(1/8)^2}\approx 7.6908$, $a_l=\frac{24\sqrt{2}}{\pi^{3/4}}\approx 14.383$, $c_l=12$, $d_l=\frac{3}{\sqrt{2}}\pi^{3/4}\approx 5.0057$.}
   
\label{tab:p(theta)}
\end{table*}

\begin{table*}
\caption{Different equations used for the non-dimensional Yarkovsky pressure $p_{\theta,\mathrm{sph}}$ throughout the article.}
\begin{center}
\renewcommand{\arraystretch}{3}
\begin{tabular}{p{2cm}|c|c|p{10cm}} 
\hline\textbf{}
Name & Equation & Range & Comments \\
\hline
linear & \raisebox{3pt}{$\displaystyle\frac{\theta}{a'_l+c'_l\theta+d'_l\theta^2}$} & $\theta\gg 1$ & analytical solution, Eq. (\ref{p_big_theta}) \\ 
\hline

perturbative asymptotics & \raisebox{-3pt}{$\displaystyle\frac{\theta}{a'_p}$} & $\theta\ll 1$ & analytical solution, Eq. (\ref{p_small_theta}) \\
\hline

unified & \raisebox{3pt}{$\displaystyle\frac{\theta}{a'_p+b'_p\theta^{1/4}+c'_l\theta+d'_l\theta^2}$} & all $\theta$ & analytical equation constructed to fit both asymptotics, Eq. (\ref{p_unified_analytic}) \\ 
\hline

corrected equatorial fit~6 & \raisebox{-7pt}{$\begin{aligned}\frac{m\cdot\theta^{-\mu} + l + n\cdot\theta^{\nu}}{\theta^{-\mu} + k + \theta^{\nu}}\times \\
\frac{\theta}{a'_p+x\theta^{\xi}+y\theta^{\eta}+z\theta^{\zeta}+d'_l\theta^2}
\end{aligned}$} & all $\theta$ & 6-parametric fit to equatorial element corrected by Eq. (\ref{knee}); $x=6.9138$, $\xi=0.25260$, $y=8.7966$, $\eta=0.66496$, $z=5.8895$, $\zeta=1.1225$, $m=0.9089$, $\mu=0.2737$, $n=1.1250$, $\nu=0.8726$, $l=0.3333$, $k=0.3309$\\ 
\hline

fit 6 & $\displaystyle\frac{\theta}{a'_p+x\theta^{\xi}+y\theta^{\eta}+z\theta^{\zeta}+d'_l\theta^2}$ & all $\theta$ & fit to numerical results; $x=6.9314$, $\xi=0.26193$, $y=8.7402$, $\eta=0.71822$, $z=5.4702$, $\zeta=1.1313$\\ 
\hline
\end{tabular}
\end{center}
{\footnotesize Note: The equations use the following notations: $f_p=\frac{3\sqrt{\pi}\Gamma(13/8)}{4\Gamma(17/8)}=1.1250$, $f_l=\frac{3\sqrt{\pi}\Gamma(19/8)}{4\Gamma(23/8)}=0.90887$, $a'_p=-\frac{45\sqrt{\pi}\Gamma(-3/8)}{4\sqrt{2}\Gamma(1/8)f_p}\approx 6.3640$, $b'_p=\frac{675\Gamma(-3/8)^2}{16\sqrt{2}\Gamma(1/8)^2 f_p}\approx 6.8365$, $a'_l=\frac{24\sqrt{2}}{\pi^{3/4}f_p}\approx 12.786$, $c'_l=\frac{12}{f_l}=13.203$, $d'_l=\frac{3\pi^{3/4}}{\sqrt{2}f_l}\approx 5.5077$.}
   
\label{tab:p_sph(theta)}
\end{table*}

\end{document}